\documentclass[11pt]{iopart}
\pdfoutput=1
\usepackage[utf8]{inputenc}
\usepackage[english]{babel}
\usepackage[T1]{fontenc}
\usepackage{amsmath, bbold, enumerate,enumitem,xcolor, amssymb,amsthm}
\usepackage{hyperref}
\usepackage[sort&compress,numbers,square,comma]{natbib}
\usepackage{mathtools}
\usepackage{tikz}
\usepackage{lipsum}
\usepackage{wrapfig}
\usepackage{tablefootnote}
\usepackage{booktabs}
\usepackage{capt-of}
\usepackage[normalem]{ulem}
\usepackage{geometry}
\usepackage{float}
\usepackage[most]{tcolorbox}
\usepackage{tikz}
\usepackage{varwidth}
\usepackage{wrapfig}
\usepackage{braket}
\usepackage[normalem]{ulem}
\usepackage{tcolorbox}
\tcbuselibrary{breakable}
\tcbset{enhanced,colback=red!5!white,colframe=red!65!black,fonttitle=\bfseries, boxrule=0.7pt}
\newtcolorbox{mybox}[1]{colback=red!5!white,colframe=red!65!black,fonttitle=\bfseries,title=#1,boxrule=0.7pt}
\usepackage{makecell}
\newcolumntype{L}[1]{>{\raggedright\let\newline\\\arraybackslash\hspace{0pt}}p{#1}}
\setenumerate{topsep=5pt,itemsep=0ex,partopsep=1ex,parsep=1ex}

\begin{document}

%%%%%%%%%%%%%%%%%%%%%%%%%%%%%%%%%%%%%%%%%%%%%%%%%%%%%%%%%%%%%%%%%%%%%%%%%%%%%%%%%%%%%%%%

\title{Extensive entanglement between coupled Tomonaga-Luttinger liquids in and out of equilibrium}

\author{Taufiq Murtadho $^1$, Marek Gluza $^1$, Nelly H. Y. Ng $^{1,2}$}
\address{$^{1}$ School of Physical and Mathematical Sciences, Nanyang Technological University, 637371, Singapore}
\address{$^{2}$ Centre for Quantum Technologies, Nanyang Technological University Singapore, 50 Nanyang Avenue, 639798, Singapore.}
\eads{\mailto{fiqmurtadho@gmail.com}, \mailto{nelly.ng@ntu.edu.sg}}

%%%%%%%%%%%%%%%%%%%%%%%%%%%%%%%%%%%%%%%%%%%%%%%%%%%%%%%%%%%%%%%%%%%%%%%%%%%%%%%%%%%%%%%%
%%%%%%%%%%%%%%%%%%%%%%%%%%%%%%%%%%%%%%%%%%%%%%%%%%%%%%%%%%%%%%%%%%%%%%%%%%%%%%%%%%%%%%%%
%%%%%%%%%%%%%%%%%%%%%%%%%%%%%%%%%%%%%%%%%%%%%%%%%%%%%%%%%%%%%%%%%%%%%%%%%%%%%%%%%%%%%%%%

\begin{abstract}
    Quantum entanglement exists in nature but is absent in classical physics, hence it fundamentally distinguishes quantum from classical theories. 
    While entanglement is routinely observed for few-body systems, it is significantly more challenging to witness in quantum many-body systems. Here, we theoretically study entanglement \textit{between} two parallel and spatially separated Tomonaga-Luttinger liquids (TLLs) partitioned \textit{along} the longitudinal axis. In particular, we focus on 1D Bose gases as a realization of TLLs and investigate two experimentally relevant situations: tunnel-coupled gases at finite temperatures and after coherent splitting.
    In both scenarios, we analytically calculate the logarithmic negativity and identify a threshold temperature below which the system is entangled. Notably, this threshold temperature is accessible in the current and near-term 1D Bose gas experiments.
    Furthermore, we investigate the crossover between quantum and classical correlations in the vicinity of the threshold temperature by comparing logarithmic negativity with mutual information. 
    We argue that the initial mutual information established by the coherent splitting is conserved in TLL dynamics, thus preventing certain generalized Gibbs ensembles from being reached during prethermalization. Moreover, both logarithmic negativity and mutual information are found to scale extensively with the subsystem's length.
    Although the ground-state entanglement between coupled TLLs has been predicted to be extensive, this setting is largely overlooked compared to other partitions. 
    Our work extends the study of entanglement between coupled TLLs to finite temperatures and out-of-equilibrium regimes, and provides a strategy towards experimental detection of extensive entanglement in quantum many-body systems at finite temperatures.
\end{abstract}

%%%%%%%%%%%%%%%%%%%%%%%%%%%%%%%%%%%%%%%%%%%%%%%%%%%%%%%%%%%%%%%%%%%%%%%%%%%%%%%%%%%%%%%%
\vspace{-10cm}
\maketitle
\markboth{}{}
%%%%%%%%%%%%%%%%%%%%%%%%%%%%%%%%%%%%%%%%%%%%%%%%%%%%%%%%%%%%%%%%%%%%%%%%%%%%%%%%%%%%%%%%

%%%%%%%%%%%%%%%%%%%%%%%%%%%%%%%%%%%%%%%%%%%%%%%%%%%%%%%%%%%%%%%%%%%%%%%%%%%%%%%%%%%%%%%%
%\tableofcontents
\section{Introduction}\label{sec:intro}
%%%%%%%%%%%%%%%%%%%%%%%%%%%%%%%%%%%%%%%%%%%%%%%%%%%%%%%%%%%%%%%%%%%%%%%%%%%%%%%%%%%%%%%%

Entanglement is a type of quantum correlation that has no analogue in classical physics, hence it serves as a fundamental signature to distinguish quantum from classical systems \cite{bell1964einstein}. In addition to its fundamental importance, entanglement is also a key resource for quantum information processing \cite{horodecki2009quantum}. Entanglement has been routinely observed and utilized in well-controlled few-body systems \cite{nicolai2018observation,friis2019entanglement}, but detecting and characterizing entanglement in quantum many-body systems, in particular those described by effective quantum field theories (QFTs), is challenging. Such many body-systems involve many degrees of freedom and therefore have a potential to host a macroscopic amount of entanglement. At the same time, experimental access is typically limited to coarse-grained observables. The immensity of the Hilbert space versus the limited number of measurement settings makes it difficult to experimentally witness entanglement~\cite{raeisi2011coarse, wang2013precision}, especially at finite temperatures \cite{anders2008thermal, anders2008entanglement}. 

For many-body systems constrained to one-dimension (1D), their low-energy behaviors are effectively described by the Tomonaga-Luttinger liquid (TLL) \cite{giamarchi2003quantum}. When two such TLLs are arranged in parallel and coupled through tunneling, they form a quantum simulator for the sine-Gordon model \cite{gritsev2007linear, schweigler2017experimental, schweigler2021decay}—a paradigmatic example of a strongly interacting integrable field theory important for various physical phenomena of interest \cite{cuevas2014sine}. This model is experimentally realized in parallel 1D Bose gases on an atom chip, where thousands of ultracold atoms are trapped in an elongated potential with a double-well transverse cross section \cite{schumm2005matter, hofferberth2008probing,hofferberth2007non,gring2012relaxation, schweigler2017experimental, rauer2018recurrences, tajik2023verification, schweigler2021decay}. 
Advances in coherent control \cite{hofferberth2006radiofrequency,tajik2019designing, kuriatnikov2025fastcoherentsplittingboseeinstein} and Gaussian tomography \cite{gluza2020quantum, gluza2022mechanisms} for this setup have allowed experiments to probe quantum information in and out of equilibrium \cite{tajik2023verification, stefan2025experimentally}, and yet experimental study of entanglement in this setup is scarce.
%observing entanglement for this setup is still elusive. 
\begin{figure}
    \centering
    \includegraphics[width=\linewidth]{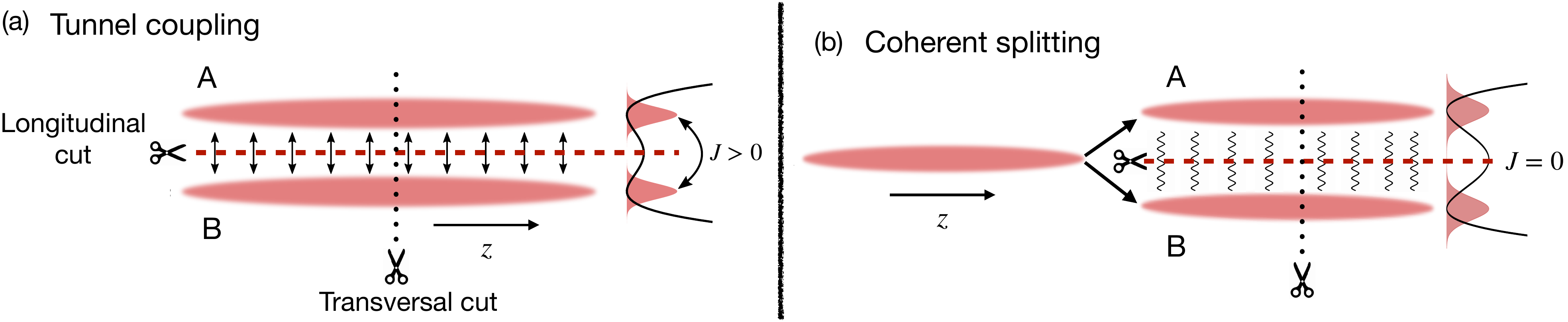}
    \caption{Two parallel 1D quasicondensates $A$ and $B$, extended along the $z$-axis, each with its own phase and density fluctuations. While typically the system is partitioned across the $z$-axis (black dotted line, transversal cut), we partition the system along the $z$-axis (red dashed line, longitudinal cut). (a) Coupling between $A$ and $B$ is introduced in the Hamiltonian through tunneling, whose strength is tunable by adjusting the height of the double-well barrier. (b) Initial correlation between $A$ and $B$ is dynamically induced by coherently splitting the trap from a single to double-well.}
    \label{fig:setup_and_partition}
\end{figure}

One of the main bottlenecks is that most results in the literature of entanglement for 1D systems focus on partitioning across the 1D axis (also known as the transversal cut, see Fig. \ref{fig:setup_and_partition})~\cite{vidal2003entanglement, nicolas2006boundary, calabrese2008spectrum, calabrese2009entanglement, eisert2010colloquium, calabrese2011the}, 
%\nn{[just these two?]} 
dividing the total length of the system into non-overlapping segments. It is a ubiquitous expectation that the ground state entanglement entropy for this partition follows the area law in the presence of a spectral gap \cite{eisert2010colloquium}, which in 1D implies that the entanglement does not grow with longer subsystem length. For a gapless system the entanglement grows with subsystem's length albeit only logarithmically~\cite{calabrese2009entanglement}. While parallel 1D Bose gases experiments confirm the area law for mutual information at finite temperatures \cite{tajik2023verification}, entanglement with transversal partition has not been reported, possibly because only microscopic amounts can be naturally generated and it is quickly diluted by thermal noise \cite{tajik2023verification,raeisi2011coarse, wang2013precision,anders2008thermal, anders2008entanglement}. 

In this work, we highlight a better strategy to observe entanglement, i.e. across the \textit{longitudinal cut} (dashed line in Fig. \ref{fig:setup_and_partition}), which partitions the system along the 1D axis and divides it into two transversely separated 1D gases. This partitioning is the more natural strategy for experimentally observing entanglement, since the mechanisms for generating strong correlations are well understood. For example, correlation can be introduced via tunnel-coupling interaction, implemented by adjusting the height of the double-well barrier (Fig. \ref{fig:setup_and_partition}a) \cite{tajik2023verification,schweigler2017experimental,rauer2018recurrences, schweigler2021decay}, or by coherently splitting the potential from a single to double-well (Fig. \ref{fig:setup_and_partition}b) \cite{gring2012relaxation, hofferberth2007non,kuhnert2013multimode,langen2013local,langen2015experimental, smith2013prethermalization}. Entanglement between spatially separated (3D) Bose–Einstein condensates (BEC) has been demonstrated \cite{lange2018entanglement,fadel2018spatial,colciaghi2023einstein} using a witness based on the total spin of each atomic cloud. In parallel 1D quasicondensates, a similar method has observed quantum-limited number squeezing \cite{max2024quantum, zhang2024squeezing} which witness particle-entanglement \cite{sorensen2001many, fadel2020relating}. Meanwhile, extensive entanglement between finite-momentum modes of \textit{distinct components} in BEC has been theoretically predicted in the ground state \cite{yoshino2021intercomponent}, but its experimental demonstration at finite temperature regimes probed in the laboratory is still missing. 
In parallel 1D quasicondensates, the measurement of these finite-momentum excitations is accessible through spatially resolved matter-wave interferometry \cite{van2018projective, murtadho2025systematic, murtadho2025measurement}, 
providing opportunity for probing entanglement in the longitudinal partition.

Beyond its experimental relevance, analyzing the longitudinal cut is also of theoretical interest. 
It enables quantum simulation of 
`field space entanglement' \cite{mollabashi2014entanglement, teresi2011entanglementquantumfields,mozaffar2016entanglement, nakai2017entanglement, mohammadi2024capacity}, i.e. entanglement between multiple quantum fields coexisting and interacting in a common (1+1D) spacetime, which is not as well understood as the more conventional spatial partition of a single quantum field. 
In condensed matter physics, this setting is also relevant for studying `bulk entanglement' in systems with ladder geometries \cite{li2024numerical,poilblanc2010entanglement, launchli2012entanglement, chen2013quantum} and electron-phonon entanglement \cite{roosz2021entanglement, Roosz2022densitymatrix}. Previous theoretical work in this direction \cite{furukawa2011entanglement, lundgreen2013entanglement, yoshino2021intercomponent, roosz2021entanglement} has shown that the ground state entanglement entropy between two coupled TLLs scales linearly with subsystem's length, and therefore extensive with respect to subsystem's size.
Yet, this extensive entanglement is spatially structured since it arises from local interactions, distinguishing them from generic many-body excited states obeying volume law scaling but with a high degree of scrambling \cite{miao2021eigenstate, bianchi2022volume}.

To our knowledge, the entanglement between two parallel coupled TLLs has not been characterized beyond the ground state. We address this gap by investigating entanglement and mutual information in realistic experimental scenarios: at finite temperatures and after coherent splitting. We specify the conditions and the parameter regimes where entanglement can be observed and show that these are achievable in the current and near-term experiments.

The paper is organized as follows. After this introduction, we present preliminaries in Sec. \ref{sec:prelim} which include the Hamiltonian of coupled 1D Bose gases (Subsec. \ref{subsec:coupled_1d_gas}), covariance matrix for Gaussian fields (Subsec. \ref{subsec:cov_mat}), and procedures to compute entanglement and mutual information of mixed Gaussian states (Subsec. \ref{subsec:ent_mi}). In Sec. \ref{sec:thermal_ent_mi}, we present our results on logarithmic negativity and mutual information of tunnel-coupled system at finite temperatures. Then, in Sec. \ref{sec:split_ent_and_mi} we extend our analysis to a non-equilibrium state by considering the system after coherent splitting. Finally, we present the summary and outlook in Sec.
\ref{sec:summ_and_outlook}. 

\section{Preliminaries} \label{sec:prelim}
\subsection{Coupled 1D Bose gases Hamiltonian}\label{subsec:coupled_1d_gas}

We consider two parallel 1D Bose gases with repulsive contact interaction, extended along the $z$-axis which we will refer to as the \emph{longitudinal} direction. Each 1D gas is described by a bosonic field operator $\hat{\psi}^a(z)$ with superscript $a,b\in \{A,B\}$ labeling the two gases (Fig. \ref{fig:setup_and_partition}). We will use a small letter in the superscript to represent an abstract label while a capital letter is reserved for a specific subsystem. At low energies, the fluctuation of the field operator can be decomposed into phase $\hat{\phi}^a(z)$ and density fluctuations $\delta\hat{n}^a(z)$ via the Madelung transformation
\begin{equation}
\hat{\psi}^{a}(z) =e^{i\hat{\phi}^a(z)}\sqrt{n_{\rm 1D} +\delta \hat{n}^{a}(z)},
\label{eq:Madelung}
\end{equation}
with $n_{\rm 1D}$ being the mean density. For simplicity, we assume identical box potential for both gases, giving a uniform and equal mean density. The phase and density fluctuations are conjugate variables satisfying canonical commutation relation
\begin{align}
[\delta \hat{n}^{a}(z),\hat{\phi}^{b}(z')]&=i\delta(z-z')\delta_{ab},
\label{eq:canonical}
\end{align}
with $\delta(z-z^\prime)$ being understood as a smeared $\delta$-function within the healing length $\xi_h = \hbar/\sqrt{mgn_{\rm 1D}}$, below which the low-energy theory is expected to break down \cite{mora2003extension}. Here $\hbar$ is the reduced Planck constant, $m$ is the atomic mass, and $g$ is the interatomic repulsion strength. 

The microscopic Hamiltonian of the system is given by the Lieb-Liniger model \cite{lieb1963exact}. At low-energies, it is well described by the Tomonaga-Luttinger liquid (TLL) Hamiltonian with quantum pressure correction (see \ref{appdxA}) %\sout{\cite{schweigler2019correlations}}
\begin{equation}
\hat{H}^{a}_{\rm TLL}= \int_{0}^{L}dz\left[\frac{\hbar^{2} n_{\rm 1D}}{2m}(\partial_{z}\hat{\phi}^{a}(z))^{2}+\frac{g}{2}(\delta \hat{n}^{a}(z))^{2}+\frac{\hbar^{2}}{8m n_{\rm 1D}} (\partial_z\delta\hat{n}^a)^2\right].
\label{eq:TLL_Hamiltonian}
\end{equation} 
Here, we have assumed explicit forms of the speed of sound and the Luttinger parameter valid in the quasicondensate regime. The last term is the quantum pressure correction included to take into account deviation from the low-energy linear dispersion.

We are interested in the entanglement between the two TLLs. A common way to introduce correlation between them is to add a tunnel-coupling of the form $-\hbar J(\hat{\psi}^{A\dagger}\hat{\psi}^B+\text{h.c.})$ to the Hamiltonian with $J$ being the tunneling strength. Such tunneling induces a non-linear interaction of the form
\begin{equation}
    \hat{H} = \hat{H}_{\rm TLL}^A+\hat{H}_{TLL}^B-2\hbar Jn_{\rm 1D} \int_{0}^{L} \cos \left[\hat{\phi}^A(z) - \hat{\phi}^B(z)\right]\; dz
    \label{eq:tunnel_coupled_hamiltonian}
\end{equation}
in the low-energy model, assuming small density fluctuations $\delta\hat{n}^a(z)\ll n_{\rm 1D}$. This interaction can be experimentally realized by adjusting the height of the double-well barrier separating the two quasicondensates.

The coupled Hamiltonian \eqref{eq:tunnel_coupled_hamiltonian} is usually treated by expressing the fields in the symmetric ($+$) and antisymmetric ($-$) basis
\begin{align}
\hat{\phi}^{\pm}(z)&=\frac{1}{\sqrt{2}}\left[\hat{\phi}^{A}(z)\pm \hat{\phi}^{B}(z)\right] \qquad \qquad
\delta\hat{n}^{\pm}(z) =\frac{1}{\sqrt{2}}\left[\delta\hat{n}^{A}(z)\pm \delta\hat{n}^{B}(z)\right],
\label{eq:plus_minus_basis_transform}
\end{align}
which preserves the canonical commutation relation.
We emphasize that the antisymmetric (relative) and symmetric (common) phases $\hat{\phi}^{\pm}(z)$ commute and can be measured simultaneously at single-shot level in experiments \cite{murtadho2025measurement, van2018projective, murtadho2025systematic}, thereby providing access to the individual phase profiles $\hat{\phi}^{A,B}(z)$ up to a constant. In the $\pm$ basis, the Hamiltonian decouples into Luttinger liquid in the symmetric sector and a sine-Gordon model in the antisymmetric sector \cite{gritsev2007linear, schweigler2017experimental}
\begin{align}
\hat{H} &= \hat{H}_{\rm TLL}^+ +\hat{H}_{\rm TLL}^{-} -2\hbar J n_{\rm 1D}\int_{0}^{L} \cos \left[\sqrt{2}\hat{\phi}^-(z)\right]\; dz\nonumber\\
&\approx \hat{H}_{\rm TLL}^{+} +\hat{H}^-_{\rm TLL} +2\hbar J n_{\rm 1D}\int_{0}^{L} \left(\hat{\phi}^-(z)\right)^2\; dz.
\label{eq:KG_hamiltonian}
\end{align}
In Eq. \eqref{eq:KG_hamiltonian}, we have used a harmonic approximation for the cosine term valid in the strong coupling regime (when $J$ dominates over other couplings), giving a Gaussian model for the coupled systems.
In this paper, we focus on this Gaussian limit where tunneling only adds a mass term to the antisymmetric sector, realizing the Klein-Gordon field \cite{tajik2023experimental, tajik2023verification,gluza2022mechanisms}.

Up to second-order in the phase and density fields, the Hamiltonian can be diagonalized using a Bogoliubov transformation, resulting in two-species bosonic theory comprising a massless mode ($+$) and a massive ($-$) mode (see \ref{appdxA}). The diagonalized Hamiltonian of the non-zero momentum modes is given by
\begin{equation}
    \hat{H} \approx \sum_{k \neq 0} \varepsilon_k^+(\hat{b}_k^+)^\dagger \hat{b}_k^+ + \varepsilon_k^-(\hat{b}_{k}^-)^\dagger \hat{b}_{k}^-,
    \label{eq:diagonalized_hamiltonian}
\end{equation}
where $k$ are the momentum of the quasiparticles, and $\hat{b}_k^{\pm}$ are the associated bosonic annihilation operators. For a free dispersion of the form $E_k = (\hbar k)^2/2m$, the Bogoliubov energy spectrum of the massless ($\varepsilon_k^+$) and massive  ($\varepsilon_k^-$) modes are \cite{mora2003extension, petrov2003bose, whitlock2003relative}
\begin{equation}
    \varepsilon_k^+ = \sqrt{E_k(E_k +2gn_{\rm 1D})} \qquad \varepsilon_k^- = \sqrt{(E_k+2\hbar J)(E_k+2\hbar J +2gn_{\rm 1D})}.
    \label{eq:bogoliubov_spectra}
\end{equation}
Note that Eq. \eqref{eq:bogoliubov_spectra} takes into account deviation from linear dispersion in the phononic regime. If the system is partitioned into the symmetric ($+$) and antisymmetric ($-$) fields, as usually done in the literature, there would be no correlation between the two sectors in thermal equilibrium. But our interest lies in the longitudinal cut shown in Fig. \ref{fig:setup_and_partition}, where the system is partitioned into gas $A$ and gas $B$. With this choice, the fields remain correlated at thermal equilibrium as long as $J>0$. Even without any tunneling ($J=0$), correlation can be generated in the initial state by pushing the system far away from equilibrium, e.g., by coherently splitting the trap from a single to double-well \cite{schumm2005matter}. 

\subsection{Covariance matrix for Gaussian fields}\label{subsec:cov_mat}
We characterize the state of the system by the covariance matrix $\Gamma$ in momentum space, which is a sufficient description for Gaussian fields \cite{serafini2023quantum}. To make $\Gamma$ finite dimensional, infrared (IR) and ultraviolet (UV) cutoffs must be introduced. The IR cutoff is automatically imposed by assuming finite length $L$ and choosing boundary conditions. For the latter, we choose Neumann boundary conditions, consistent with our assumption of an idealized box potential. The UV cutoff is defined by the inverse healing length $\sim \xi_h^{-1}$. We then decompose the fields into summation of discrete modes 
\begin{equation}
    \hat{\phi}^{a}(z) = \frac{1}{\sqrt{L}} \hat{\phi}_0^a+\sqrt{\frac{2}{L}} \sum_{k>0}^{k_{\Lambda}} \hat{\phi}_k^{a} \cos(kz)
    \label{eq:cosine_decomposition_1}
\end{equation}
\begin{equation}
        \delta \hat{n}^{a}(z) = \frac{1}{\sqrt{L}} \delta\hat{n}_0^a+\sqrt{\frac{2}{L}} \sum_{k>0}^{k_{\Lambda}} \delta \hat{n}^a_{k} \cos(kz),
        \label{eq:cosine_decomposition_2}
\end{equation}
where the allowed momenta are $k = N\pi / L$ with $N = 1, 2, 3, \ldots, \Lambda$ and the positive integer $\Lambda$ is chosen such that $k_{\Lambda} = \Lambda\pi / L \sim \xi_h^{-1}$. The prefactor $\sqrt{2/L}$ ensures proper commutation relation $[\delta\hat{n}^a_{k}, \hat{\phi}^b_{q}] = i \delta_{kq} \delta_{ab}$.

From here onwards, we focus on correlation in the non-zero momentum modes, e.g. we consider the case where the zero-modes $\hat{\phi}_0^a, \delta\hat{n}_0^a$ are removed from the data by phase referencing and adjusting for atom number fluctuations. This differentiates our approach from previous studies of particle-entanglement in 1D quasicondensates \cite{zhang2024squeezing, berrada2013integrated}, which relies on probing entanglement through zero-mode squeezing in the relative sector. The relevant covariance matrix $\Gamma \in \mathbb{R}^{4\Lambda \times 4\Lambda}$ is then constructed from the two-point correlation functions of the quadrature operator $\hat{Q}^a$ defined as
\begin{equation}
    \hat{Q}^a \coloneqq \begin{pmatrix}
        \delta\hat{n}^a_{1} & \cdots & \delta\hat{n}^a_{\Lambda} & 
        \hat{\phi}^a_{1} & \cdots & \hat{\phi}^a_{\Lambda}
    \end{pmatrix}^{\top}.
\end{equation}
In our case, it is convenient to organize $\Gamma$ into a $2 \times 2$ block matrix
\begin{equation}
    \Gamma = \begin{pmatrix}
        \Gamma^{AA} & \Gamma^{AB} \\
        \Gamma^{BA} & \Gamma^{BB}
    \end{pmatrix},
\end{equation}
where $\Gamma^{ab} \in\mathbb{R}^{2\Lambda \times 2\Lambda}$ are the submatrices with $(j,l)$ element given by
\begin{equation}
     \Gamma^{ab}_{jl} = \frac{1}{2} \langle \hat{Q}^a_{j} \hat{Q}^b_{l} + \hat{Q}^a_{l} \hat{Q}^b_{j} \rangle.
\end{equation}
The canonical commutation relation imposes the Heisenberg uncertainty principle on physical covariance matrices, i.e.,
\begin{equation}
    \Gamma + \frac{i}{2} (\Omega \oplus \Omega) \geq 0, \qquad \text{with} \qquad \Omega = \begin{pmatrix}
        0 & \mathcal{I}_{\Lambda} \\
        -\mathcal{I}_{\Lambda} & 0
    \end{pmatrix}
\end{equation}
being the symplectic matrix and $\mathcal{I}_\Lambda$ is the $\Lambda \times \Lambda$ identity matrix. Finally, we remark that in experiments, $\Gamma$ can be reconstructed from real space phase profiles through a Gaussian tomography scheme \cite{gluza2020quantum,gluza2022mechanisms,tajik2023verification, stefan2025experimentally}.

\subsection{Entanglement measure and mutual information for Gaussian states}\label{subsec:ent_mi}
The subsystems $A$ and $B$ are correlated as long as $\Gamma^{AB} \neq 0$, which can be due to entanglement or classical correlation. If the state of the joint system is pure, the entanglement entropy (i.e. the von Neumann entropy of the reduced density matrix) quantifies the entanglement. For a Gaussian state, it can be computed through the reduced covariance matrix $\Gamma^{aa}$ by~\cite{eisert2010colloquium, serafini2023quantum,tajik2023verification}
\begin{equation}
    S(\Gamma^{aa}) = \sum_{q=1}^\Lambda\left(\gamma_q^a+\frac{1}{2}\right)\log \left(\gamma^a_{q}+\frac{1}{2}\right)-\left(\gamma^a_{q}-\frac{1}{2}\right)\log \left(\gamma^a_{q}-\frac{1}{2}\right),
    \label{eq:ent_entropy}
\end{equation}
where $\gamma^a_{q}$ are the symplectic eigenvalues of $\Gamma^{aa}$, i.e. they are the positive eigenvalues of $i\Omega\Gamma^{aa}$, with $q\in \{1,2,..., \Lambda\}$ labeling the symplectic eigenvalues.

For mixed joint states, entanglement entropy is no longer a valid entanglement measure. Nevertheless, the positive partial transpose (PPT) criterion remains a necessary condition for separability \cite{serafini2023quantum}. For continuous variable (CV) systems, partial transposition is equivalent to flipping the sign of the momentum-like fields \cite{simon2000peres}, e.g. in our case $\hat{\phi}^B_{k}\rightarrow -\hat{\phi}^B_{k}$ for all modes $k$. Let us denote the partially transposed covariance matrix as $\Gamma^{\top_B}$ and $\gamma_q^{\top_B}$ ($q = 1,2,...,2\Lambda$) as its symplectic eigenvalues. The subsystems $A$ and $B$ are entangled if the logarithmic negativity
\begin{equation}
    E_{\mathcal{N}} \coloneqq \sum_{q=1}^{2\Lambda}\max\left\{0, -\log \left(2\gamma_q^{\top_B}\right)\right\}> 0
    \label{eq:log_neg_def}
\end{equation}
is positive. When this is true, $E_{\mathcal{N}}$ quantifies the amount of entanglement \cite{plenio2005log}. The converse is also true for two-mode Gaussian states; if $E_{\mathcal{N}} = 0$, the system is separable \cite{simon2000peres, duan2000inseparability}. But in general multimode case, $E_{\mathcal{N}}> 0$ is only a sufficient condition, i.e. there exist bound entangled states with $E_{\mathcal{N}} = 0$ \cite{werner2001bound}.  

Another useful measure of correlation is the mutual information $I(A:B)$, which quantifies the total correlations between $A$ and $B$, including both entanglement and classical correlations. For Gaussian states, it is given by
\begin{equation}
    I(A:B)  = S(\Gamma^{AA}) +S(\Gamma^{BB}) - S(\Gamma)\ . 
    \label{eq:mutual_info_definition}
\end{equation}
The first two terms can be calculated with Eq. \eqref{eq:ent_entropy}. For pure joint states, the joint entropy $S(\Gamma) = 0$, and so the mutual information is precisely twice the entanglement entropy. For mixed joint states, $S(\Gamma)$ does not vanish in general, but it is computable through Eq. \eqref{eq:ent_entropy} with the symplectic matrix being $\Omega \oplus \Omega$ and $q$ runs until $2\Lambda$. Logarithmic negativity and mutual information are the two correlation measures we will focus on in this paper. 

\section{Results: Entanglement and mutual information at finite temperatures}\label{sec:thermal_ent_mi}

Since the low-energy Hamiltonian separates into symmetric ($+$) and antisymmetric sectors ($-$), the system is described by a two-species bosonic theory comprising a massless symmetric mode and a massive antisymmetric mode. Assuming Neumann boundary conditions, we use Bogoliubov transformation \cite{petrov2003bose, whitlock2003relative, mora2003extension, rauer2018recurrences} to expand the phase and density fluctuations in terms of bosonic creation and annihilation operators $\hat{b}_k^{\pm}$ in the two sectors 
\begin{equation}
    \hat{\phi}_{k}^+ = \frac{1}{\sqrt{4n_{\rm 1D}}}\sqrt{\frac{\varepsilon_{k}^+}{E_k}}\left[\hat{b}_{k}^++\text{h.c.}\right] \qquad \delta\hat{n}_k^+ = \sqrt{n_{\rm 1D}}\sqrt{\frac{E_k}{\varepsilon_k^+}}\left[i\hat{b}_k^++\text{h.c.}\right]
    \label{eq:plus_fields}
\end{equation}
\begin{equation}
    \hat{\phi}_{k}^- = \frac{1}{\sqrt{4n_{\rm 1D}}}\sqrt{\frac{\varepsilon_{k}^-}{E_k+2\hbar J}}\left[\hat{b}_{k}^-+\text{h.c.}\right] \qquad \delta\hat{n}_k^- = \sqrt{n_{\rm 1D}}\sqrt{\frac{E_k+2\hbar J}{\varepsilon_k^-}}\left[i\hat{b}_k^-+\text{h.c.}\right],
    \label{eq:minus_fields}
\end{equation}
where $E_k = (\hbar k)^2/(2m)$ is the free particle dispersion, $\varepsilon_k^\pm$ are the Bogoliubov energy spectra defined in Eq. \eqref{eq:bogoliubov_spectra}. In this section, we will assume the system is at thermal equilibrium, having a density matrix $\rho \propto  \exp(-\beta \hat{H)}$ with inverse temperature $\beta$. The elements of the covariance matrix $\Gamma^{ab}$ can be calculated from Eqs. \eqref{eq:plus_fields} and \eqref{eq:minus_fields}. For example, the phase-phase correlations are
\begin{equation}
    \braket{\hat{\phi}^a_k \hat{\phi}_q^b} = \begin{cases}
    \frac{1}{2}\left(\braket{\hat{\phi}^+_k \hat{\phi}_k^+}+\braket{\hat{\phi}^-_k \hat{\phi}_k^-}\right)\delta_{kq} \quad \text{if} \; a = b\\
     \frac{1}{2}\left(\braket{\hat{\phi}^+_k \hat{\phi}_k^+}-\braket{\hat{\phi}^-_k \hat{\phi}_k^-}\right)\delta_{kq}\quad \text{if} \; a \neq b,
    \end{cases}
\end{equation}
and a similar equation also holds for the density-density correlations, while the phase-density correlations vanish in equilibrium. 

Crucially, all correlation functions vanish for different momentum modes $k \neq q$, implying that the $k$-mode of gas $A$ can only be correlated with the same $k$-mode of gas $B$. Characterizing entanglement thus reduces to solving a series of independent two-mode entanglement problem, for which PPT criterion is necessary and sufficient \cite{duan2000inseparability, simon2000peres}. Another important consequence is that the entanglement and mutual information are monotone under adding and removing any mode $k$. Hence, we can justifiably ignore the contribution from the zero mode $(k = 0)$ and the modes above the UV cutoff $(k>k_{\Lambda})$. The calculated logarithmic negativity and mutual information by ignoring these modes can be treated as lower bound for the true logarithmic negativity and mutual information of the system.

The transformation from $A,B$ to $\pm$ basis diagonalizes the covariance matrix while preserving the symplectic eigenvalues (see  \ref{appdxB}). This allows for analytic evaluation of the symplectic eigenvalues, and consequently also the logarithmic negativity and the mutual information. We begin by presenting our result for the logarithmic negativity
\begin{equation}
    E_{\mathcal{N}} = \sum_{k>0}^{k_{\Lambda}}\max\left\{0, -\log\left[\sqrt{C_{k,J}(1+2\eta_{k}^+)(1+2\eta_{k}^-)}\right]\right\},
    \label{eq:log_neg}
\end{equation}
where the spectral ratio $C_{k,J}$ is defined as
\begin{equation}
    C_{k,J} \coloneqq \frac{\varepsilon_k^-}{\varepsilon_k^+}\frac{E_k}{E_k+2\hbar J},
    \label{eq:spectral_ratio}
\end{equation}
and 
\begin{equation}
    \eta_{k}^{\pm} \coloneqq \braket{(\hat{b}_k^\pm)^\dagger \hat{b_k^\pm}} = [\exp(\beta\varepsilon_k^\pm)-1]^{-1}
\end{equation}
are the mean thermal occupations in the symmetric and antisymmetric sectors. The result can be generalized to the case where each sector has a different temperature $(\beta^+ \neq \beta^-)$ by appropriately defining $\eta_k^\pm$. From Eq. \eqref{eq:log_neg}, we see a competition between tunneling strength $J$ and the temperature $T = (k_B\beta)^{-1}$ in establishing entanglement ($k_B$ is the Boltzmann constant). On the one hand, tunneling tends to create entanglement since the spectral ratio satisfies $C_{k,J}\leq 1$ with the inequality being saturated at $J = 0$. This implies that at zero temperatures ($\eta_k^+ = \eta_k^- = 0$), all modes are entangled as long as $J>0$. However, as expected, temperature tends to destroy entanglement with the term $(1+2\eta_k^+)(1+2\eta_k^-)\geq 1$. 

For a fixed tunneling strength, we calculate the threshold temperature $T^*$ above which entanglement vanishes, i.e. $E_{\mathcal{N}}(T) = 0$ for all $T>T^*$ where
\begin{equation}
T^* = \sup_{0<k\leq k_{\Lambda}}\left\{ T_k^*\Big|\tanh\left(\frac{\varepsilon_k^+}{2k_BT^*_k}\right)\tanh\left(\frac{\varepsilon_k^-}{2k_BT_k^*}\right) = C_{k,J}\right\}, 
\label{eq:threshold_temp}
\end{equation}
easily solvable by root-finding. At this point, we remark that while the Hamiltonian \eqref{eq:KG_hamiltonian} can be written as coupled quantum harmonic oscillators, whose thermal entanglement and threshold temperature have been studied in a general setting \cite{anders2008entanglement, anders2008thermal}, we identify a solvable and experimentally relevant instance of such models. We also go further by evaluating the experimental attainability of the threshold temperature and exploring the entanglement scaling with tunable system parameters such as tunneling strength, system's length, and mean density. Later in the section, we will probe the transition point between quantum and classical correlation by comparing logarithmic negativity with mutual information. 
\begin{figure}
    \centering
    \includegraphics[width = 0.7\textwidth]{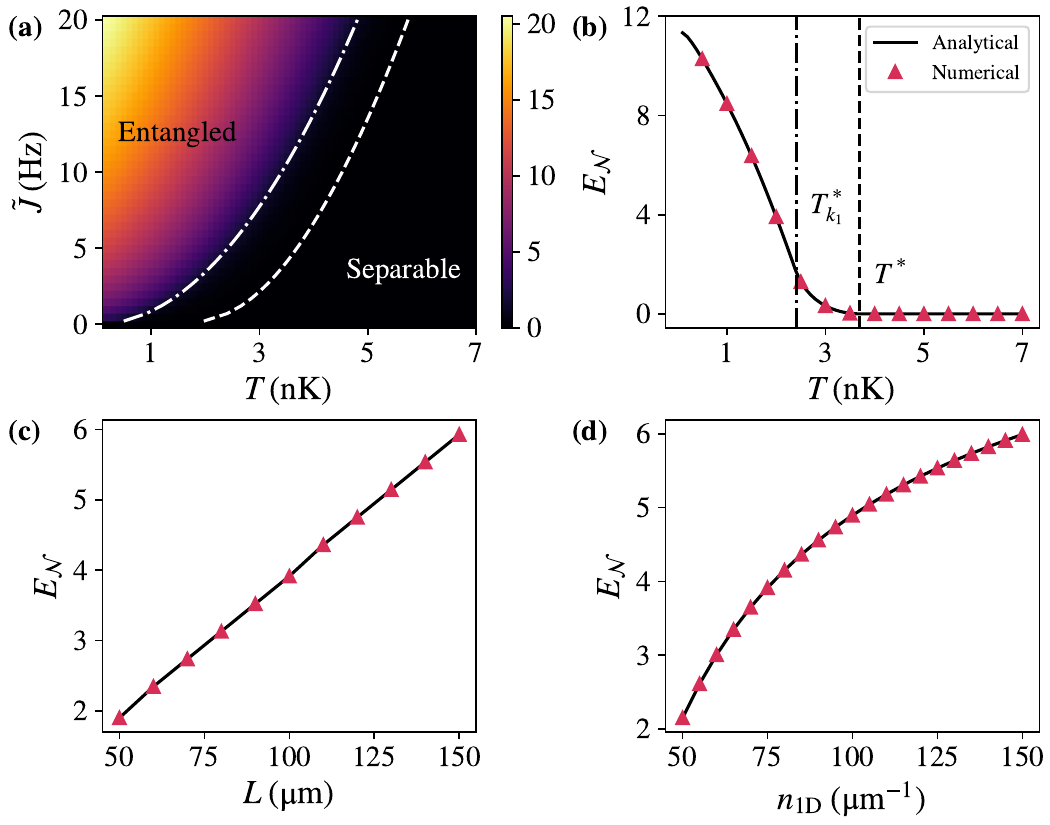}
    \caption{\textit{Entanglement at finite temperatures for tunnel-coupled gases.— } (a) Entanglement, measured by logarithmic negativity $E_{\mathcal{N}}$, as a function of tunneling strength $\Tilde{J} = J/(2\pi)$ and temperature $T$. The dashed line is the threshold temperature $T^*$ above which entanglement vanishes ($E_{\mathcal{N}} = 0$) for all modes $k \leq k_{\Lambda}$. We choose $\Lambda = 31$, which for condensate length $L = 100\; \rm \mu m$ corresponds to a healing length of $\xi_h\approx 1\; \rm \mu m$, close to the shortest length scale that can be probed in the current experiments. The dashed-dotted line is the first threshold temperature $T^*_{k_1}$ above which the entanglement in the first mode $k_1 = \pi/L$ vanishes, while still possibly present in higher momentum modes with $k>k_1$. (b) The decay of $E_{\mathcal{N}}$ with increasing $T$ for fixed tunneling $\Tilde{J} = 5 \; \rm Hz$. For panels (b)-(d), the solid lines are calculated with our analytical formula \eqref{eq:log_neg}, while the triangles are computed with numerical diagonalization. The dashed and dashed-dotted lines have the same meaning as in (a). (c) Linear scaling of $E_{\mathcal{N}} \propto L$ for fixed temperature $T = 2\; \rm nK$ and fixed tunneling strength $\Tilde{J} = 5\; \rm Hz$. The cutoff $\Lambda$ changes for different $L$ since we set $k_{\Lambda} = \Lambda\pi/L \sim \xi_h^{-1}$. In panels (a) - (c), mean density is fixed at $n_{\rm 1D } = 75 \; \rm \mu m^{-1}$, but the results do not significantly change for other $n_{\rm 1D}$ values. (d) Sub-linear scaling of $E_{\mathcal{N}}$ with mean density $n_{\rm 1D}$. The difference between scaling behavior in (c) and (d) underscores the collective nature of the entangled 1D excitations. 
    For panels (a), (b), and (d), the condensate length is fixed at $L = 100\; \rm \mu m$. The mass $m$ is the atomic mass ${}^{87}$Rb. The interaction strength $g$ is chosen to be the value reported in the experiments (see e.g. Refs.~\cite{tajik2023verification, stefan2025experimentally} among others) and fixed throughout the paper, i.e. $g \approx 2\hbar a_s\omega_\perp$ where $a_s \approx 5.2\; \rm nm$ is the scattering length, and $\omega_\perp = 2\pi \times 1.4\; \rm kHz$ is the transverse trapping frequency.}
    \label{fig:log_neg_thermal}
\end{figure}

We plot entanglement as measured by $E_{\mathcal{N}}$ with varying temperature $T$ and tunneling strength $J$ in Fig. \ref{fig:log_neg_thermal}a. We identify the parameter regimes where the system is entangled ($E_{\mathcal{N}}>0$) and where it is separable ($E_{\mathcal{N}} = 0$), separated by the threshold temperature $T^*$ curve. In the $E_{\mathcal{N}}>0$ regime, entanglement intuitively increases with stronger tunneling and drops with rising temperature. The threshold temperature $T^*$ is also found to monotonically increase with stronger tunneling. For the experimental parameters used in Refs. ~~\cite{rauer2018recurrences,  tajik2023verification, stefan2025experimentally}, $T^*$ lies in the range of $1 \sim  4\; \rm nK$, about one order of magnitude lower than the temperatures achieved in these experiments. The results do not significantly change if we allow for different temperatures in symmetric and antisymmetric sectors ($T^+ \neq T^-$), both temperatures need to be of order $1\sim 4\; \rm nK$ to observe entanglement.

For a fixed tunneling strength, the largest $T_k^*$ typically occurs for the mode with the highest accessible momentum $k_{\Lambda}$. We interpret this as the entanglement in higher momentum modes being more robust to thermal noise. As one raises the temperature, thermal noise corrupts the entanglement in all modes, but it corrupts the entanglement in low momentum modes more quickly. For instance, above the first threshold $T>T^*_{k_1}$ ($k_1 = \pi/L$), the entanglement between $A$ and $B$ vanishes for $k_1$, while still possibly present for $k>k_1$ (Fig. \ref{fig:log_neg_thermal}b). This process continues until all entanglement in the modes $k\leq k_{\Lambda}$ vanishes at $T = T^*$ and we no longer observe entanglement. Naturally, the ability to detect entanglement depends on the cutoff $\Lambda$. For a fixed cutoff $k_\Lambda \sim \xi_h^{-1}$ the system can be regarded as `effectively separable' at $T>T^*$, even though entanglement may still persist above the cutoff. Conversely, the experimental imaging resolution is typically lower than $\xi_h^{-1}$, and so depending on the specific resolution, the threshold temperature in experiments can be lower than in Fig. \ref{fig:log_neg_thermal}. 

We next investigate the scaling of entanglement with the subsystem's length $L$. Previous theoretical work \cite{furukawa2011entanglement,lundgreen2013entanglement, yoshino2021intercomponent} has shown that entanglement behaves extensively in the ground state, meaning it grows linearly with $L$. We obtain the same result for finite temperature states as long as the temperature is sufficiently far away from $T^*$ as shown in Fig. \ref{fig:log_neg_thermal}c. We next provide a simple argument to explain this. Given a fixed resolution, the number of allowed $k$-modes increases linearly with $L$ i.e. in the continuum limit $\sum_k \rightarrow (L/\pi)\int dk$. In our setting, this is the dominant mechanism by which entanglement grows with length $E_{\mathcal{N}}\propto L$. 
 %\sout{We emphasize that this extensivity arises due to the embedding of the 1D systems in 2D space, so it does not violate the standard area law for a strictly 1D system.} 
 Interestingly, the scaling behavior is sublinear if we increase mean density $n_{\rm 1D}$ instead of $L$ (Fig. \ref{fig:log_neg_thermal}d), underscoring the collective nature of excitations, as entanglement does not just scale linearly with atom number $n_{\rm 1D}L$. It also highlights the importance of interatomic interaction for observing the entanglement since $n_{\rm 1D}$ appears in Eq. \eqref{eq:log_neg} only through the interaction energy $2gn_{\rm 1D}$ in the Bogoliubov spectra $\varepsilon_k^\pm$ [Eq. \eqref{eq:bogoliubov_spectra}].

Above the threshold temperature $T>T^*$, $A$ and $B$ are effectively separable but they can still be classically correlated. We probe the transition from quantum to classical correlation by calculating the mutual information (see \ref{appdxB} for derivation)
\begin{align}
    I(A:B) = &\sum_{k>0}^{k_{\Lambda}}2\left[\left(\lambda_{k,J}+\frac{1}{2}\right) \log \left(\lambda_{k,J}+\frac{1}{2}\right) - \left(\lambda_{k,J}-\frac{1}{2}\right) \log \left(\lambda_{k,J}-\frac{1}{2}\right)\right] \nonumber \\
    &- \sum_{a = \pm} \sum_{k>0}^{k_{\Lambda}}\left[ \left(\eta_k^a+1\right) \log \left(\eta_{k}^a+1\right) - \eta_k^a \log \eta_k^a\right],
    \label{eq:thermal_mutual_info}
\end{align}
where $\lambda_{k,J}$ are the symplectic eigenvalues of the reduced covariance matrix given by
\begin{equation}
    \lambda_{k,J} = \frac{1}{4}\sqrt{\left[(1+2\eta_k^+)+C_{k,J}(1+2\eta_k^-)\right]\left[(1+2\eta_k^+)+C_{k,J}^{-1}(1+2\eta_k^-)\right]},
    \label{eq:lambdak_thermal}
\end{equation}
and $C_{k,J}$ is the spectral ratio defined in Eq. \eqref{eq:spectral_ratio}. Note that the mutual information in Eq. \eqref{eq:thermal_mutual_info} reduces to twice the entanglement entropy in the limit of zero temperature ($\eta_k^+ = \eta_k^- = 0$).

 \begin{figure}
    \centering
    \includegraphics[width=\textwidth]{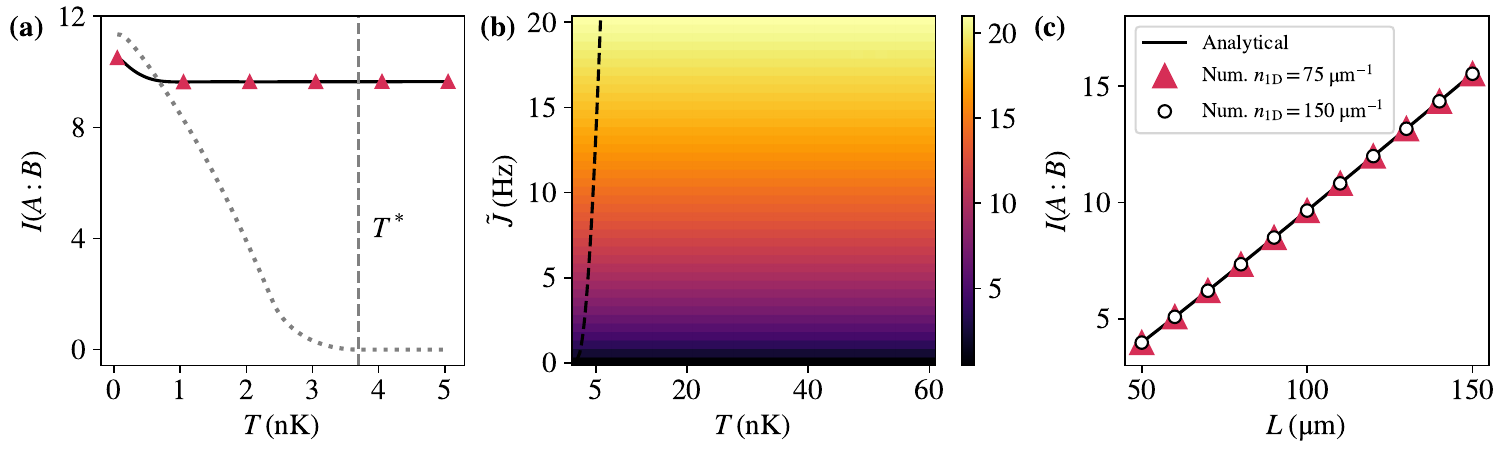}
    \caption{\textit{Mutual information at finite temperatures for tunnel-coupled gases.—} (a) Crossover from quantum to classical correlations near the threshold temperature $T^*$ (dashed line). The mutual information $I(A:B)$ (solid line) is constant while the logarithmic negativity $E_{\mathcal{N}}$ (dotted) line decays to zero. The tunneling strength is fixed at $\tilde{J} = J/(2\pi) = 5 \; \rm Hz$. (b) $I(A:B)$ tend to increase with stronger $J$ but it is almost constant with $T$, even way above $T^*$ (black dashed line). For panels (a) and (b), the condensate length is fixed at $L = 100\; \rm \mu m$. (c) Linear scaling of mutual information with system's length $I(A:B) \propto L$ for fixed temperature $T = 30\; \rm nK$ and fixed tunneling strength $\Tilde{J} = 5\; \rm Hz$. For panels (a) and (c), the solid line is calculated with Eq. \eqref{eq:thermal_mutual_info} while the triangles and the circles are calculated with numerical diagonalization. The mean density is fixed at $n_{\rm 1D} = 75\; \rm \mu m^{-1}$ in all panels except for the circles in panel (c) where $n_{\rm 1D} = 150\; \rm \mu m^{-1}$, indicating that mutual information does not significantly scale with $n_{\rm 1D}$.}
    \label{fig:thermal_mutual_info}
\end{figure}

We probe the behavior of mutual information $I(A:B)$ in the vicinity of the threshold temperature $T^*$ in Fig. \ref{fig:thermal_mutual_info}a. In contrast to decaying entanglement, $I(A:B)$ stays approximately constant. Meanwhile, the magnitude of $I(A:B)$ is comparable to $E_{\mathcal{N}}$ below $T^*$, meaning a significant portion of the correlation consists of quantum entanglement. Thus, as one approaches $T^*$, the entanglement must get transformed into classical correlations to keep the mutual information constant. 

The approximate independence of mutual information with respect to temperature persists far above $T^*$ as shown in Fig. \ref{fig:thermal_mutual_info}b. Intuitively, this can be explained if the role of temperature is only to rescale the magnitude of the local fluctuations in $A$ and $B$ without affecting their shared information. Instead, the shared information is governed by the tunneling strength, as supported by $I(A:B)$ monotonically increasing with stronger tunneling. Finally, for the same reason as the logarithmic negativity, the mutual information also scales linearly with the subsystem's length $I(A:B)\propto L$ (Fig. \ref{fig:thermal_mutual_info}c). But unlike logarithmic negativity, our analytical and numerical results suggest that the mutual information does not significantly scale with mean density. 

So far, the correlation between the subsystems $A$ and $B$ is induced by tunnel-coupling at the Hamiltonian level and investigated in thermal equilibrium. In the context of experiments in Refs. \cite{gring2012relaxation,schweigler2017experimental, rauer2018recurrences, schweigler2021decay, tajik2023verification}, the threshold temperature for observing entanglement between the two gases is an order of magnitude lower than what has been experimentally demonstrated. In the next section, we explore another scenario where the threshold temperature is within experimental reach. The scenario involves pushing the system far away from equilibrium by coherent splitting.
\section{Results: Entanglement and mutual information after coherent splitting}\label{sec:split_ent_and_mi}

In the coherent splitting scenario, the system is assumed to be initialized in thermal equilibrium of a single well at $t<0$. Then, the single-well potential is coherently split into a double-well at $t = 0$ within a timescale much shorter than the interaction timescale ($\Delta t_{\rm split}\ll \xi_h/c$ with $c = \sqrt{gn_{\rm 1D}/m}$ being the speed of sound), such that its effect is analogous to spatially extended beam splitters. Let $\hat{b}_k^0$ be the phonon annihilation operator before the splitting (for momentum $k$), and $\hat{b}^{A,B}_k$ be the phonon annihilation operators after the splitting, respectively. The case of coherent splitting is analogous to an equal beam splitter with one input being $\hat{b}^0_k$, occupied by a thermal state, and the other input being an ancilla $\hat{b}^{\rm anc}_k$, occupied by a vacuum. At the output (i.e. after the splitting), all the initial thermal energy gets transferred to the symmetric sector due to the beam splitter operation $\hat{b}^+_k = (\hat{b}^A_k + \hat{b}^B_k)/\sqrt{2} = \hat{b}^0_k$. On the other hand, the antisymmetric modes $\hat{b}^{-}_k = (\hat{b}^A_k -\hat{b}^B_k)/\sqrt{2} = \hat{b}^{\rm anc}_k$ have minimum uncertainties as they are occupied by the vacuum of the ancilla. If we also reasonably assume that the input $\hat{b}^0_k$ and $b^{\rm anc}_k$ are initially uncorrelated, it follows that $\hat{b}^+_k$ and $\hat{b}^-_k$ are uncorrelated at the output.

Meanwhile, locally the atoms undergo a random process of going either to the left or right well. This random process leads to a binomial density distribution in each small region (of size $\sim \xi_h$), giving noise to the local relative density $\delta\hat{n}^-(z)$ proportional to the mean density $n_{\rm 1D}$, while reducing fluctuation in the relative phase $\hat{\phi}^-(z)$ up to the minimum uncertainty relation. Following this argument, the momentum correlation in the antisymmetric sector after the instantaneous splitting is phenomenologically modeled as \cite{gring2012relaxation, geiger2014local, foini2015non}
\begin{equation}
\braket{\hat{\phi}_{k}^-\hat{\phi}_{q}^-}\Big|_{t = 0}= \frac{1}{4n_{\rm 1D}}\frac{\delta_{kq}}{r^2} \qquad \qquad\braket{\delta\hat{n}_{k}^-\delta\hat{n}_{q}^-}\Big|_{t = 0} = n_{\rm 1D}r^2\delta_{kq},
\label{eq:init_corr}
\end{equation}
while the phase-density correlations vanish (see \ref{appdxC} for a more theoretically systematic derivation). Here, $r$ is the number squeezing parameter, previously used as one of the key global parameters for probing squeezing in 1D gases (denoted $\xi_N$ in Refs.~\cite{zhang2024squeezing, berrada2013integrated, kuriatnikov2025fastcoherentsplittingboseeinstein, geiger2014local}). The symmetric fluctuation is modeled as thermal with a temperature equal to the condensate before the splitting, and uncorrelated with those in the antisymmetric sector. The case where the system is initialized in the ground state before the splitting can be recovered by setting temperature to zero. We remark that 
the precise control and characterization of the initial correlation after coherent splitting is a subject of ongoing research \cite{michael2019from, kuriatnikov2025fastcoherentsplittingboseeinstein, van2021josephson}.

In Subsec. \ref{subsec:init_corr} we will study the logarithmic negativity and the mutual information immediately after the coherent splitting ($t = 0$). Then, in Subsec. \ref{subsec:relaxation}, we discuss how the conservation of mutual information constrains relaxation toward a prethermalized state ($t>0$) \cite{gring2012relaxation, geiger2014local, langen2013local}. 

\subsection{Initial entanglement and mutual information at $t = 0$}\label{subsec:init_corr}

For the initial non-equilibrium state after coherent splitting, %we assume
the symmetric and antisymmetric sectors are separated and there is no correlation between modes with different momenta. These features also appear in the equilibrium tunnel-coupled case (Sec. \ref{sec:thermal_ent_mi}), allowing us to analytically calculate the logarithmic negativity and the mutual information. We thus extend the calculation done in Sec. \ref{sec:thermal_ent_mi} to this non-equilibrium state. The details of the calculation is given in \ref{appdxD}. Here, we present the expression for logarithmic negativity

\begin{equation}
        E_{\mathcal{N}} = \sum_{k}\max\left\{0, -\log\left(\sqrt{\mathcal{C}_{k,r}(1+2\eta_k^+)}\right)\right\}+\max\left\{0, -\log\left(\sqrt{\mathcal{C}_{k,r}^{-1}(1+2\eta_k^+)}\right)\right\},
        \label{eq:log_neg_split}
\end{equation}
with a squeezed spectral factor
\begin{equation}
    \mathcal{C}_{k,r} \coloneqq \frac{r^2\varepsilon_k^+}{E_k}.
    \label{eq:squeezed_spectral}
\end{equation}
From Eq. \eqref{eq:log_neg_split}, we see again a competing behavior: squeezing $r$ and the interaction energy $gn_{\rm 1D}$ (appearing through $\varepsilon_k^+$) generate entanglement, while temperature appearing through $\eta_k^+$ suppresses it. The first term dominates for squeezing parameters $r \geq 1$, whereas the second term may dominate for $r \ll 1$. 

\begin{figure}
    \centering
\includegraphics[width=\textwidth]{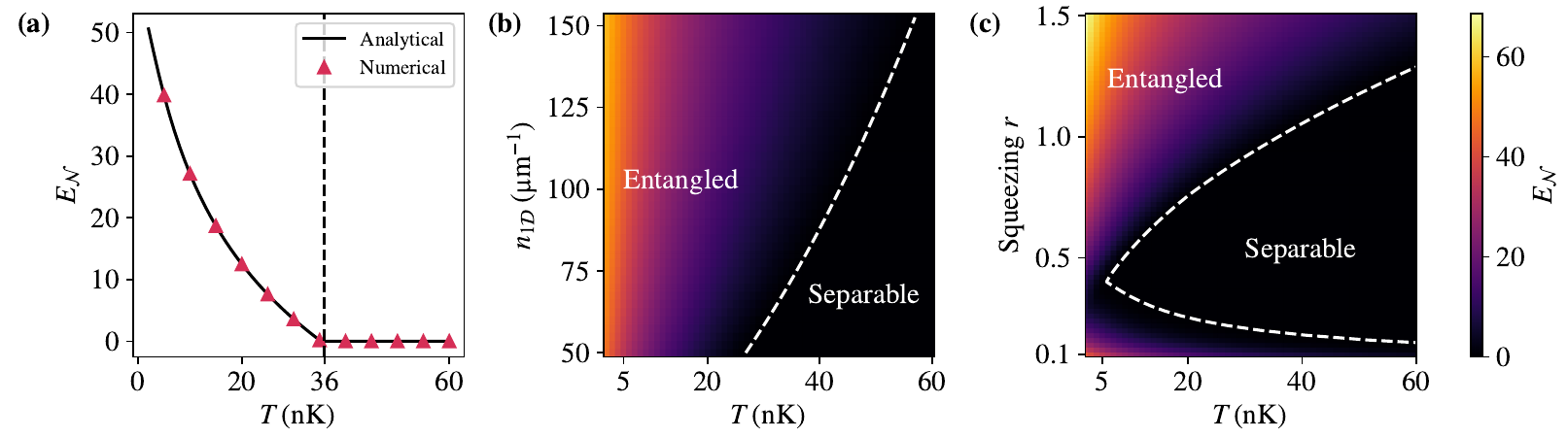}
    \caption{\textit{Entanglement immediately after coherent splitting}.— (a) Logarithmic negativity $E_{\mathcal{N}}$ as a function of temperature $T$ for $r = 1$ and $n_{\rm 1D} = 75\; \rm \mu m^{-1}$. All dashed lines indicate the threshold temperature $T^*$. For parameters in panel (a) the threshold temperature is calculated to be $T^*\approx 36\; \rm nK$. The solid line is calculated with Eq. \eqref{eq:log_neg_split} and the triangles are computed with numerical diagonalization. Entanglement as a function of temperature and (b) mean density $n_{\rm 1D}$ (fixed $r = 1$), and (c) squeezing parameter $0.1\leq r\leq 1.5$ (fixed $n_{\rm 1D} = 75\; \rm \mu m^{-1}$). The obtained range for $T^*$ is accessible in the present day experiments \cite{kuhnert2013multimode, tajik2023verification}. Both $E_{\mathcal{N}}$ and $T^*$  behave non-monotonically with $r$. The condensate length is fixed at $L = 100\; \rm \mu m$ for all panels, while further numerics show that logarithmic negativity still scales linearly with length $E_{\mathcal{N}} \propto L$, as long as the temperature is sufficiently far away from the threshold temperature.}
    \label{fig:coh_split_entanglement}
\end{figure}

Surprisingly, the threshold temperature $T^*$ can be expressed in a closed form,
\begin{equation}
    T^* = \sup_{0<k\leq k_{\Lambda}}\left\{ T_{k}^* = \frac{\varepsilon_k^+}{k_B}\left(\ln \left[\sqrt{\frac{\mathcal{C}_{k,r}+\mathcal{C}_{k,r}^{-1}+2}{\mathcal{C}_{k,r}+\mathcal{C}_{k,r}^{-1}-2}}\right]\right)^{-1}\right\}.
    \label{eq:split_threshold_temp}
\end{equation}
Note that $\mathcal{C}_{k,r}+\mathcal{C}_{k,r}^{-1}>2$ so that $T_{k}^*$ is always well-defined.
In Fig. \ref{fig:coh_split_entanglement}, we plot $E_{\mathcal{N}}$ as a function of $T$ for realistic experimental parameters. We first focus on fixed squeezing $r = 1$ and mean density $n_{\rm 1D} = 75\; \rm \mu m^{-1}$ (Fig. \ref{fig:coh_split_entanglement}a) and calculate the threshold temperature to be
$T^* \approx 36\; \rm nK$. The threshold temperature can be raised further by increasing interaction energy $gn_{\rm 1D}$ (Fig. \ref{fig:coh_split_entanglement}b) or by increasing squeezing parameter $r$ (Fig. \ref{fig:coh_split_entanglement}c). For realistic range of experimental parameters, $T^*$ is found to be between $30 \sim 60\; \rm nK$, which is readily accessible in the present day experiments \cite{tajik2023verification, kuhnert2013multimode}. 

Moreover, we find that both $E_{\mathcal{N}}$ and $T^*$ behave non-monotonically with the squeezing parameter $r$ (Fig. \ref{fig:coh_split_entanglement}c). The system exhibits entanglement for both large ($r > 1$) and small ($r < 0.4$) squeezing, but the strength and the regime of entanglement are strongly suppressed at intermediate squeezing values ($0.4 < r < 1$). This can be understood from the initial correlation in Eq. \eqref{eq:init_corr}. As the source of the entanglement is the strong correlation between the two phase fields (suppression of the relative phase fluctuation), setting $r> 1$ ($r<1$) would enhance (suppress) this correlation. But when $r$ is very small (e.g. $r<0.4$), we are in the opposite regime where the density fields are strongly correlated (relative density fluctuation is suppressed) and so entanglement re-emerge through the second term in Eq. \eqref{eq:log_neg_split}. 

To gain more insight into the crossover between quantum and classical correlations after coherent splitting, we compute the mutual information (see \ref{appdxD} for derivation)
\begin{align}
    I(A:B) = & \sum_{k>0}^{k_{\Lambda}}2\left[\left(\lambda_{k,r} +\frac{1}{2}\right)\log\left(\lambda_{k,r}+\frac{1}{2}\right) - \left(\lambda_{k,r} - \frac{1}{2}\right)\log\left(\lambda_{k,r} - \frac{1}{2}\right)\right] \nonumber\\
    & -(1+\eta_k^+)\log(1+\eta_k^+)+\eta_k^+\log\eta_k^+,
    \label{eq:split_mutual_info}
\end{align}
where
\begin{equation}
    \lambda_{k,r} = \frac{1}{4}\sqrt{\left(1+2\eta_k^++\mathcal{C}_{k,r}\right)\left(1+2\eta_k^++\mathcal{C}_{k,r}^{-1}\right)}.
    \label{eq:reduced_symp_coh}
\end{equation}
In the case where the system is prepared in the ground state before the splitting, entanglement entropy can be calculated from Eq. \eqref{eq:split_mutual_info} and Eq. \eqref{eq:reduced_symp_coh} by setting $\eta_k^+ = 0$. Note that the $\lambda_{k,r}$ above has a similar form as that of Eq. \eqref{eq:lambdak_thermal} since in both scenarios the covariance matrices have the same structure, differing only in the matrix elements in the antisymmetric sector.

\begin{figure}
    \centering
    \includegraphics[width=\textwidth]{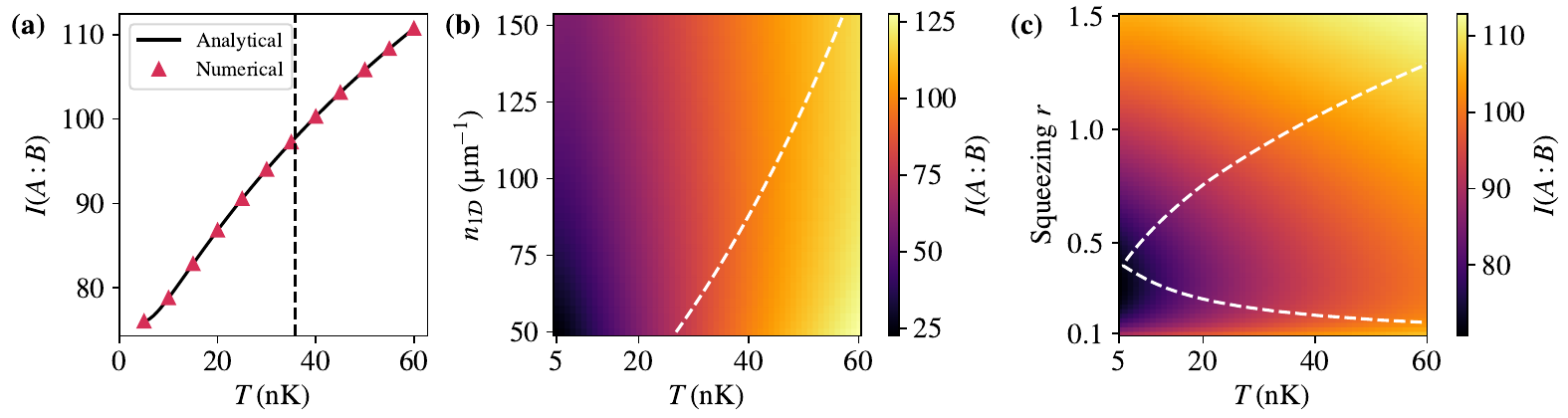}
    \caption{\textit{Mutual information immediately after coherent splitting}.— (a) The increase of mutual information $I(A:B)$ with increasing temperature $T$ for fixed squeezing $r = 1$ and mean density $n_{\rm 1D} = 75\; \rm \mu m^{-1}$, in contrast with decaying $E_{\mathcal{N}}$ in Fig. \ref{fig:coh_split_entanglement}a. The solid line is calculated with Eq. \eqref{eq:split_mutual_info} and the triangles are computed with numerical diagonalization. (b) The mutual information $I(A:B)$ varies slowly with mean density $n_{\rm 1D}$. Here the squeezing is fixed at $r = 1$. (c) Similar to entanglement, $I(A:B)$ also changes non-monotonically with squeezing parameter $r$ ($n_{\rm 1D} = 75\; \rm \mu m^{-1}$). In all panels, the dashed lines represent the threshold temperature $T^*$, same as in Fig. \ref{fig:coh_split_entanglement}, and the condensate length is fixed at $L = 100\; \rm \mu m$, while further numerics show that mutual information still scales linearly with length $I(A:B) \propto L$.}
    \label{fig:coh_split_mutual_info}
\end{figure}

In Fig. \ref{fig:coh_split_mutual_info}a, we plot the mutual information $I(A:B)$ as a function of temperature $T$ for fixed squeezing $r = 1$ and fixed mean density $n_{\rm 1D} = 75\; \rm \mu m^{-1}$. Contrary to the decay of entanglement (Fig. \ref{fig:coh_split_entanglement}a), $I(A:B)$ grows with increasing $T$, irrespective of the threshold. The same growth with temperature is also observed for different values of the mean density and the squeezing parameter (Fig. \ref{fig:coh_split_mutual_info}b - c). This behavior follows from Eq. \eqref{eq:init_corr}: since the relative sector is initially dominated by quantum noise, which is small compared to thermal fluctuations, the phase fluctuations in A and B are nearly identical after the splitting. As the temperature rises, these shared fluctuations become larger, leading to an increase in the mutual information. Thus, the splitting produces both quantum and classical correlations, with the classical correlations dominating as the temperature increases.
Similar to entanglement, $I(A:B)$ is also found to behave non-monotonically with the squeezing parameter. In Fig. \ref{fig:coh_split_mutual_info}c, we observe that $I(A:B)$ is strongly suppressed at low temperatures ($T<20\; \rm nK$) and intermediate squeezing regime ($0.2<r<1$), where entanglement is also suppressed (Fig. \ref{fig:coh_split_entanglement}c).

\subsection{Relaxation constrained by mutual information conservation}\label{subsec:relaxation}

In experiments \cite{gring2012relaxation, langen2013local}, the double-well barrier after coherent splitting is set to be high enough such that no tunneling is allowed between the two gases $A$ and $B$. Consequently, the system's dynamics follows that of decoupled TLL [Eq. \eqref{eq:tunnel_coupled_hamiltonian} with $J = 0$], i.e. $\hat{H} = \hat{H}^+_{\rm TLL}+\hat{H}^-_{\rm TLL}$. Interestingly, the relative phase fluctuation $\hat{\phi}^-(z)$ is experimentally observed to relax into a thermal-like metastable state with an effective inverse temperature $\beta^{-} = (g n_{\rm 1D}/2)^{-1}$, independent of the system’s initial temperature before the splitting~\cite{gring2012relaxation, langen2015experimental, smith2013prethermalization}. The temperature in the relative sector is inferred from the measured relative phase correlation functions and its full distribution functions \cite{gring2012relaxation, smith2013prethermalization, kitagawa2011dynamics, hofferberth2008probing}, whose behavior is found to be consistent with thermal ensemble after relaxation. This phenomenon is an example of prethermalization, widespread in out-of-equilibrium integrable systems \cite{berges2004prethermalization, mori2018thermalization}. The prethermalization in these experiments is modeled by multimode dephasing due to coherent rotation between the relative density and the relative phase fluctuations \cite{gring2012relaxation, gluza2022mechanisms, kuhnert2013multimode}.

On the contrary, here we argue that local unitary dynamics generated by decoupled TLL \textit{cannot} bring the coherently split initial state [Eq. \eqref{eq:init_corr}] to a Generalized Gibbs ensemble (GGE) \cite{rigol2007relaxation, langen2015experimental} of the form 
\begin{equation}
    \rho_{\rm GGE}= \frac{1}{Z}\exp(-\beta^-\hat{H}_{\rm TLL}^-) \otimes \exp(-\beta^+\hat{H}_{\rm TLL}^+)\ ,
    \label{eq:prethermal_GGE}
\end{equation}
which is often conjectured to be the relaxed prethermalized state in the past theoretical, numerical, and experimental studies \cite{kaminishi2018entanglement, stimming2011dephasing, gring2012relaxation, langen2015experimental}. Here we focus specifically on the GGE of the form Eq. \eqref{eq:prethermal_GGE} without ruling out other forms of GGE (with other conserved quantities). The state \eqref{eq:prethermal_GGE} cannot be reached under TLL dynamics because the Hamiltonian not only decouples in the $\pm$ basis but it is also decoupled in the $A,B$ basis: $\hat{H} = \hat{H}_{\rm TLL}^A + \hat{H}_{\rm TLL}^B$. Consequently, the unitary dynamics generated by $\hat{H}$ must conserve the mutual information (and in general, also entanglement) between subsystem $A$ and subsystem $B$, i.e., $I(A:B, t) = I(A:B, 0)$ for all $t>0$.  

Meanwhile, we can compute the mutual information for GGE states of the form \eqref{eq:prethermal_GGE} as a special case of Eq. \eqref{eq:thermal_mutual_info} with $J = 0$ and $\beta^+\neq \beta^-$. We obtain
\begin{align}
    I_{\rm GGE}(A:B) &= \sum_{k>0}^{k_{\Lambda}}2\left[(1+\overline{\eta}_k)\log (1+\overline{\eta}_k)-\overline{\eta}_k\log \overline{\eta}_k\right]\nonumber\\
     &- \sum_{a = \pm} \sum_{k>0}^{k_{\Lambda}}\left[ \left(\eta_k^a+1\right) \log \left(\eta_{k}^a+1\right) - \eta_k^a \log \eta_k^a\right], 
     \label{eq:prethermalized_mi}
\end{align}
where $\overline{\eta}_k = \frac{1}{2}(\eta_k^++\eta_k^-)$ is the average of thermal occupations in both sectors. By comparing Eq. \eqref{eq:prethermalized_mi} with Eq. \eqref{eq:split_mutual_info}, one finds that mutual information is not conserved, hence a contradiction. 

Since the symmetric sector is assumed to be always thermal, our results show that the antisymmetric sector cannot reach the Gibbs state from the given initial correlation \eqref{eq:init_corr} via TLL dynamics. Indeed, the relaxation of correlation functions including the density-density correlation \cite{foini2015non}, and cross-correlation function \cite{kaminishi2015entanglement,kaminishi2018entanglement} after a coherent split have been theoretically analyzed. It was revealed that the correlation functions are \textit{not} exactly compatible with a thermal equilibrium in the long-time limit. However, numerically, they look very similar that they may practically be indistinguishable in experiments. Our result highlights another physical constraint---the mutual information conservation---which extends beyond the behavior captured by correlation functions in the subtle process of relaxation and prethermalization in isolated 1D quantum gases.

Alternatively, the system could still prethermalize into GGE of the form \eqref{eq:prethermal_GGE} in experiments, but at least one of the following must be true: the initial correlation is not of the form Eq. \eqref{eq:init_corr}, or the system does not evolve according to the TLL model. The TLL model is only effective and the exact model almost certainly features non-zero interactions, although the time-scale at which interaction becomes relevant is unclear \cite{mazets2009dephasing, rauer2018recurrences, kuhnert2013thermalization,Cataldini_PRX.12.041032}. In the long-time limit, the system is expected to thermalize at phonon-levels into a thermal state with $\beta^+ = \beta^- = \beta$, having zero mutual information $I(A:B) = 0$ for decoupled double-well ($J = 0$). This process is expected to occur via anharmonic correction to TLL which include three-body phonon interaction \cite{burkov2007decoherence, mazets2009dephasing, stimming2011dephasing, huber2018thermalization, micheli2022phonon}. 

Introducing phonon-phonon interaction into the effective model can lead to interesting dynamics for the entanglement and mutual information studied here. This is because the evolution become effectively non-unitary as excitations may leak to momentum states above the cutoff and to the non-Gaussian degrees of freedom. In experiments, the spatial resolution of the measurements places a cut-off on the momenta that can be resolved~\cite{murtadho2025systematic,van2018projective}. A universal flow of momentum excitations has previously been argued \cite{langen2016prethermalization}. This means that the dynamics will appear to `move' certain excitations out of view while also `move' some excitations into view. Moreover, interaction may lead to a build up of non-Gaussian correlations, implying that the entropic quantities evaluated under the assumption of the state being Gaussian may not be accurate. However, as long as the non-Gaussian state does not depart too far from being Gaussian, the entropy is continuous under the assumption of a finite energy~\cite{winter2016tight}. If $\rho$ is an arbitrary state, there exists a unique Gaussian state $\rho_G$ with the same second moments $\Gamma$ as $\rho$. The entropy computed using the Gaussian formula for $\rho_G$ will overestimate the true entropy of $\rho$, i.e. $S(\rho)\le S(\rho_G)$. This is because Gaussian entropy disregards higher-order correlations as if they were unobserved, which for a non-Gaussian state corresponds to a lack of knowledge reflected in the higher entropy value. How exactly these momentum cutoff and non-Gaussian correlations affect entanglement and mutual information during relaxation dynamics is a subject for further research.

\section{Summary and Outlook}\label{sec:summ_and_outlook}

We have studied entanglement and mutual information between coupled Tomonaga-Luttinger liquids partitioned along the longitudinal axis at finite temperatures and after coherent splitting. We unambiguously established that the entanglement and mutual information across this partition are extensive, by showing that both quantities scale linearly with the subsystem length. 
Importantly, we also identified the threshold temperature for observing entanglement, and found that in the case of coherent splitting, the temperature range is already within experimental reach.

Our study explores the dependence of entanglement and mutual information on key physical parameters, including temperature, tunneling strength, mean density, and the squeezing parameter. Our analysis indicates that mutual information—a key correlation measure employed in previous experiments \cite{tajik2023verification, stefan2025experimentally}—often reflects classical correlations and therefore cannot reliably capture the entanglement features of the system in the scenarios considered here. Nevertheless, it provides us with other important insights, including the quantum-to-classical correlation transition near the threshold temperature, and the conservation of mutual information in the non-interacting TLL dynamics, which forbids prethermalization to GGE states with distinct temperatures in the symmetric and antisymmetric sectors, unless phonon-phonon interaction is taken into account. 

A strong motivation for this work is the goal to experimentally detect extensive entanglement in many-body quantum systems at finite temperatures. Our results suggest that such detection is not only feasible in 1D Bose gas experiments, but the required measurement protocol is also accessible. Specifically, from the coherent TLL evolution of the symmetric and antisymmetric phases correlation functions
\begin{equation}
    G_\pm (z,z^\prime, t) = \braket{[\hat{\phi}^{\pm}(z,t) - \hat{\phi}^{\pm}(0, t)][\hat{\phi}^{\pm}(z^\prime, t) - \phi^{\pm}(0,t)]},
\end{equation}
accessible from density interference patterns after time-of-flight,
one may apply the existing Gaussian tomography protocol \cite{gluza2020quantum,gluza2022mechanisms} to reconstruct the full covariance matrix $\Gamma$, including the density-density sectors as their information gets encoded into the phase fluctuations during the evolution. Such a tomographic protocol has already been applied in experiments to reconstruct the covariance matrix and information-theoretic measures in the antisymmetric sector \cite{tajik2023verification, stefan2025experimentally}. Reconstructing logarithmic negativity and mutual information studied in this paper involves extending this tomography protocol to both symmetric and antisymmetric sectors. This is made possible by recent developments in matter-wave interferometry after time-of-flight, allowing for single-shot extraction of not only relative but also common phase fluctuations \cite{murtadho2025measurement}. Thus, while many studies on entanglement in field theories remain largely theoretical, this work highlights a clear pathway that could be pursued experimentally.

On the other hand, experimental detection of entanglement brings about other challenges that have not been addressed here. For instance, certifying the reliability of Gaussian tomography reconstruction for witnessing entanglement includes taking into account statistical fluctuations, inherent systematic error in the phases extraction protocols \cite{murtadho2025systematic, murtadho2025measurement}, and imperfections in the imaging, which are known to introduce convolutions to the measured density interference patterns \cite{schweigler2019correlations, gluza2020quantum}. Furthermore, typical experiments involve harmonic trapping, meaning that the system deviates from being fully homogeneous. In such situations, correlations between different momentum modes can become non-negligible, so logarithmic negativity is no longer a necessary witness of entanglement. Additionally, contribution from the zero mode as well as from the modes above the UV cutoff may influence the overall correlations. Even in experiments with a box potential \cite{tajik2019designing, rauer2018recurrences, schweigler2021decay, tajik2023verification}, the mean density curvature due to imperfect box should be scrutinized. A numerical and experimental study on precise certification of entanglement and mutual information from density interference data is an important direction for future work.

When considering tunnel-coupled gases at finite temperatures, we have focused on the Klein-Gordon limit which is well-described by Gaussian theory. Note that moving away from this limit towards the sine-Gordon regime introduces non-Gaussian correlations, which lie beyond the scope of our covariance-matrix-based analysis. In this regime, our evaluation of logarithmic negativity continues to serve as a reliable witness of entanglement. However, it effectively quantifies the entanglement of a mixed Gaussian state rather than the true non-Gaussian state, which leads us to expect a suppression of the detected entanglement signal. A full characterization would in principle require access to the complete density matrix, which is not directly experimentally accessible. The systematic characterization of non-Gaussian entanglement in continuous-variable, many-body systems (e.g. via $n$-point correlation functions) remain a major open challenge for the community.

Another direction of interest to both theory and experiment is to consider relaxation dynamics of entanglement and mutual information after a sudden quench or in the presence of a generic time-dependent tunneling strength. One may theoretically study these problems in the limit of low-energy theories—the Tomonaga-Luttinger liquid and the sine-Gordon model—or probe deviations from low energy theories and investigate the impact of cutoff and non-Gaussian correlations as discussed in the previous section. The latter is more challenging theoretically but it is highly relevant for experiments. Equipped with reliable Gaussian tomography, the long-time relaxation of Gaussian entanglement and mutual information can in principle be probed experimentally \cite{kuhnert2013thermalization}. 

The presented work thus contributes to ongoing efforts in both theory and experiment for exploring entanglement in low-dimensional strongly correlated quantum matter, both in and out of equilibrium. By providing a timely theoretical foundation, this work paves the way for rigorous quantum simulation of interacting quantum field theories in cold-atomic platforms.

\section{Acknowledgements}
We are grateful to
Jörg Schmiedmayer and L\'eonce Dupays for their valuable feedback. We also thank Paola Ruggiero, Maximilian Prüfer, Giuseppe Vitagliano, Julia Math\'e, Nicky Kai Hong Li, and Tiang Bi Hong for insightful discussions. This work was supported by the start-up grant of the Nanyang Assistant Professorship of
Nanyang Technological University, Singapore, and the Tier 2 grant from Ministry of Education, Singapore.

\appendix
%\newpage
\markboth{}{}

\section{Diagonalization of tunnel-coupled 1D Bose gases Hamiltonian}\label{appdxA}
In the main text, we write down the low-energy model of tunnel-coupled 1D Bose gases Hamiltonian in terms of the (quantum-pressure corrected) Tomonaga-Luttinger liquid (TLL) and the Klein-Gordon (KG) Hamiltonian. Here, we derive the low-energy Hamiltonian from the microscopic model and use Bogoliubov theory to diagonalize it up to second-order in the phase and density fields. We start from the tunnel-coupled Lieb-Liniger (LL) model
\begin{equation}
    \hat{H} = \int dz\left[\hat{\mathcal{H}}_{\rm LL}^A +\hat{\mathcal{H}}_{\rm LL}^{B} + \hat{\mathcal{H}}^{AB}_J\right],
\end{equation}
with Hamiltonian densities
\begin{equation}
    \hat{\mathcal{H}}_{\rm LL}^a = \frac{\hbar^2}{2m}\partial_z\hat{\Psi}^{a\dagger} \partial_z\hat{\Psi}^a+ \frac{g}{2}\hat{\Psi}^{a\dagger}\hat{\Psi}^{a\dagger}\hat{\Psi}^a\hat{\Psi}^a \qquad a\in\{A,B\},
\end{equation}
\begin{equation}
    \hat{\mathcal{H}}_J^{AB} = -\hbar J[\hat{\Psi}^{A\dagger}\hat{\Psi}^B + \hat{\Psi}^{B\dagger}\hat{\Psi}^A].
\end{equation}
We then write the field operators in terms of phase and density fields as in Eq. \eqref{eq:Madelung} and expand the Hamiltonian up to the second-order. 

In this approximation, the Lieb-Liniger term $\mathcal{H}_{\rm LL}^a$ is approximately given by the TLL model with quantum pressure correction [Eq. \eqref{eq:TLL_Hamiltonian}]
\begin{equation}
    \hat{\mathcal{H}}_{\rm LL}^{a} \approx \frac{\hbar^{2} n_{\rm 1D}}{2m}(\partial_z\hat{\phi}^a)^2 + \frac{g}{2} (\delta\hat{n}^{a})^2+\frac{\hbar^{2}}{8m n_{\rm 1D}} (\partial_z\delta\hat{n}^a)^2.
\end{equation}
The tunneling Hamiltonian density $\hat{\mathcal{H}}_{J}^{AB}$ up to the second-order expansion is given by
\begin{equation}
    \hat{\mathcal{H}}_{J}^{AB} \approx -\hbar J n_{\rm 1D}\left[-\left(\hat{\phi}^A-\hat{\phi}^B\right)^2+\frac{\delta\hat{n}^A+\delta\hat{n}^B}{n_{\rm 1D}}-\frac{\left(\delta\hat{n}^A-\delta\hat{n}^B\right)^2}{4n_{\rm 1D}^2}\right].
\end{equation}
In the quasicondensate regime, the density fluctuation is suppressed $\delta\hat{n}^a(z)\ll n_{\rm 1D}$ and so the last term contribution is often neglected when writing down the low-energy Klein-Gordon model \eqref{eq:KG_hamiltonian}. But here we keep them to diagonalize the model with Bogoliubov theory, taking into account deviation from the KG spectrum. 

Transforming into symmetric and antisymmetric basis as in Eq. \eqref{eq:plus_minus_basis_transform} gives a decoupled Hamiltonian
\begin{equation}
    \hat{H} \approx \int dz\left[\mathcal{H}^++\mathcal{H}^-\right],
    \label{eq:hamiltonian}
\end{equation}
where
\begin{equation}
    \hat{\mathcal{H}}^+ = \frac{\hbar^{2} n_{\rm 1D}}{2m}(\partial_z\hat{\phi}^+)^2+ \frac{\hbar^{2}}{8m n_{\rm 1D}} (\partial_z\delta\hat{n}^+)^2 + \frac{g}{2} (\delta\hat{n}^+)^2 -\hbar J\delta\hat{n}^+,
    \label{eq:Hamiltonian_density_plus}
\end{equation}
\begin{equation}
    \hat{\mathcal{H}}^- = \frac{\hbar^{2} n_{\rm 1D}}{2m}(\partial_z\hat{\phi}^-)^2+2\hbar J(\hat{\phi}^-)^2+ \frac{\hbar^{2}}{8m n_{\rm 1D}} (\partial_z\delta\hat{n}^-)^2 + \frac{g}{2} (\delta\hat{n}^-)^2+\frac{\hbar J}{2n_{\rm 1D}}(\delta\hat{n}^-)^2.
    \label{eq:Hamiltonian_density_minus}
\end{equation}
The Hamiltonian \eqref{eq:hamiltonian} with quadratic Hamiltonian densities \eqref{eq:Hamiltonian_density_plus} and \eqref{eq:Hamiltonian_density_minus} is diagonalized by the mode decomposition \eqref{eq:cosine_decomposition_1}-\eqref{eq:cosine_decomposition_2} and canonical transformations \eqref{eq:plus_fields}-\eqref{eq:minus_fields} in the main text, with energy spectrum given by the Bogoliubov spectrum \eqref{eq:bogoliubov_spectra}.

\section{Derivation of logarithmic negativity and mutual information for tunnel-coupled gases in equilibrium}\label{appdxB}

From the expressions of the Bogoliubov decomposition of the fields in the symmetric $(+)$ and antisymmetric $(-)$ basis [Eqs. \eqref{eq:plus_fields} and \eqref{eq:minus_fields}] one can obtain analytical formula for all elements of the covariance matrix ($\Gamma$) in $A, B$ basis as follows
\begin{equation}
    \braket{\hat{\phi}_k^a\hat{\phi}_k^b} = \begin{cases}
        &\frac{1}{8n_{\rm 1D}}\left(\frac{\varepsilon_k^+}{E_k}(1+2\eta_{k}^+) +  \frac{\varepsilon_k^-}{E_k+2\hbar J}(1+2\eta_{k}^-)\right) \quad \text{if}\; a = b\\
        & \frac{1}{8n_{\rm 1D}}\left(\frac{\varepsilon_k^+}{E_k}(1+2\eta_{k}^+) -  \frac{\varepsilon_k^-}{E_k+2\hbar J}(1+2\eta_{k}^-)\right) \quad \text{if}\; a \neq b,
    \end{cases}
    \label{eq:thermal_cov_elems_1}
\end{equation}
\begin{equation}
    \braket{\delta\hat{n}_k^a\delta\hat{n}_k^b} = \begin{cases}
        &\frac{n_{\rm 1D}}{2}\left(\frac{E_k}{\varepsilon_k^+}(1+2\eta_{k}^+) +  \frac{E_k+2\hbar J}{\varepsilon_k^-}(1+2\eta_{k}^-)\right) \quad \text{if}\; a = b\\
        & \frac{n_{\rm 1D}}{2}\left(\frac{E_k}{\varepsilon_k}(1+2\eta_{k}^+) -  \frac{E_k+2\hbar J}{\varepsilon_k^-}(1+2\eta_{k}^-)\right) \quad \text{if}\; a \neq b,
        \label{eq:thermal_cov_elems_2}
    \end{cases}
\end{equation}
and all other elements are zeros. Note that the covariance matrix $\Gamma$ is non-diagonal, it has non-zero elements corresponding to $a \neq b$. 

We first derive the formula for the logarithmic negativity $E_{\mathcal{N}}$ [Eq. \eqref{eq:log_neg}]. We start by performing partial transposition $\Gamma^{\top_B}$, which in our case corresponds to the transformation $\braket{\hat{\phi}_k^A\hat{\phi}_k^B}\rightarrow -\braket{\hat{\phi}_k^A\hat{\phi}_k^B}$. Next, we need to evaluate the symplectic eigenvalues of $\Gamma^{\top_B}$ defined by the positive eigenvalues of $i(\Omega \oplus \Omega)\Gamma^{T_B}$. This can be analytically done by first applying a symplectic transformation
\begin{equation}
     \Tilde{\Gamma}^{\top_B}= \mathcal{S}\Gamma^{\top_B} \mathcal{S}^\top \qquad \text{with} \qquad 
    \mathcal{S} = \frac{1}{\sqrt{2}}\begin{pmatrix}
        I_{2\Lambda}& I_{2\Lambda}\\
        I_{2\Lambda} & -I_{2\Lambda}
    \end{pmatrix}.
    \label{eq:symplectic_transform}
\end{equation}
The transformation $\mathcal{S}$ preserves the symplectic structure $\mathcal{S}(\Omega \oplus \Omega)\mathcal{S}^T = \Omega \oplus \Omega$ and leaves the symplectic eigenvalues invariant, while the transformed matrix $\Tilde{\Gamma}^{\top_B}$ takes a diagonal form. For a diagonal covariance matrix, Schur's determinant formula allows us to write the symplectic eigenvalues as the square root of the product of the diagonal elements in each quadrature. In our context, they are given by
\begin{align}
    \nu_k^{\pm}\Big|_{\Gamma^{\top_B}}&= \sqrt{\big(\braket{\hat{\phi}_k^A\hat{\phi}_k^A}\mp\braket{\hat{\phi}_k^A\hat{\phi}_k^B} \big)\big(\braket{\delta\hat{n}_k^A\delta\hat{n}_k^A}\pm\braket{\delta\hat{n}_k^A\delta\hat{n}_k^B}\big)}\\
    & = \frac{1}{2}\sqrt{C_{k,J}^{\pm 1}(1+2\eta_{k}^+)(1+2\eta_{k}^-)},
    \label{eq:expression}
\end{align}
where $C_{k,J} = (\varepsilon_k^-/\varepsilon_k^+)(E_k/(E_k+2\hbar J))$ is the spectral ratio as defined in Eq. \eqref{eq:spectral_ratio}. 

Entanglement is detected by logarithmic negativity if $\Gamma^{\top_B}$ violates the Heisenberg uncertainty principle, i.e. either $\nu_k^+<1/2$ or $\nu_k^-<1/2$ [See Eq. \eqref{eq:log_neg_def}]. Meanwhile, one can check from Eq. \eqref{eq:expression} that $\nu_k^-\geq 1/2$ by noticing that
\begin{equation}
    C_{k,J}^{-1} = \left[\frac{\varepsilon_k^-}{\varepsilon_k^+}\frac{E_k}{E_k+2\hbar J}\right]^{-1} = \frac{(E_k+2\hbar J)(E_k+2gn_{\rm 1D})}{E_k(E_k+2\hbar J +2gn_{\rm 1D})} \geq 1, 
\end{equation}
and $\sqrt{(1+2\eta_k^+)(1+2\eta_k^-)}\geq 1$. Hence, only $\nu_k^+$ contribute to the logarithmic negativity. By substituting $\nu_k^+$ from Eq. \eqref{eq:expression} to Eq. \eqref{eq:log_neg_def} we recover Eq. \eqref{eq:log_neg} in the main text. The expression for the threshold temperature [Eq. \eqref{eq:threshold_temp}] is obtained by setting $\nu_k^+ = 1/2$.

The derivation for mutual information $I(A:B)$ [Eq. \eqref{eq:thermal_mutual_info}] proceeds in a similar way. We first compute the symplectic eigenvalues $\lambda_k$ of the diagonal reduced covariance matrix $\Gamma^{AA} = \Gamma^{BB}$. For $a \in \{A,B\}$ we find
\begin{align}
    \lambda_k\Big|_{\Gamma^{aa}} &= \sqrt{\braket{\hat{\phi}_k^a\hat{\phi}_k^a}\braket{\delta\hat{n}_k^a\delta\hat{n}_k^a}} \\
    &= \frac{1}{4}\sqrt{\left[(1+2\eta_k^+)+C_{k,J}(1+2\eta_k^-)\right]\left[(1+2\eta_k^+)+C_{k,J}^{-1}(1+2\eta_k^-)\right]},
\end{align}
Then, for calculating the symplectic spectrum of the joint covariance matrix $\Gamma$, we perform the symplectic transformation \eqref{eq:symplectic_transform} to $\Gamma$ and obtain the symplectic eigenvalues $\sigma_k^\pm$
\begin{equation}
    \sigma_k^\pm\Big|_{\Gamma}= \left(\eta_k^\pm +\frac{1}{2}\right)
\end{equation}
Using the definition of mutual information and the formula for the von-Neumann entropy [Eqs. \eqref{eq:mutual_info_definition} and \eqref{eq:ent_entropy}] we recover the result for the mutual information in Eq. \eqref{eq:thermal_mutual_info}. 

\section{Derivation of initial correlation after coherent splitting}\label{appdxC}
In this section, we derive the correlation ansatz for coherently split gases in Eq. \eqref{eq:init_corr}. Suppose that we divide the gases into small patches of size $d \sim \xi_h$ with $\xi_h$ being the healing length. Let $N$ be the number of atoms on each patch. The probability of an atom going to the well $A$ is given by $p$ (for balanced splitting $p = 1/2$). The probability of having a configuration $N_A$ atoms on well $A$ (and consequently also $N_B = N - N_A$ on well $B$) is described by a binomial distribution
\begin{equation}
    P(N,N_A, p) = \binom{N}{N_A}p^{N_A}(1-p)^{N-N_A}.
    \label{eq:binom_dist}
\end{equation}
We can write Eq. \eqref{eq:binom_dist} in a different way. Suppose that we want to know the probability of the \textit{difference} in the atomic numbers in well $A$ and $B$ being $2\Delta N$ with $\Delta N \in \{-N/2, ..,0, ..,N/2\}$ where we have assumed even $N$ for simplicity. In other words, we write
\begin{equation}
    N_A = \frac{N}{2} +\Delta N \qquad N_B = \frac{N}{2} - \Delta N.
\end{equation}
The probability for this configuration is given by
\begin{equation}
    P(N, \Delta N, p) = \binom{N}{\frac{N}{2}+\Delta N}p^{N/2+\Delta N}(1-p)^{N/2 - \Delta N}.
\end{equation}
One can compute the variance of $\Delta N$ for balanced splitting $p = 1/2$ as
\begin{equation}
     \braket{\Delta N^2} = \frac{N}{4}.
\end{equation}
We now rephrase this variance in terms of variance of the relative density field $\delta n ^-(z)$ as defined in \eqref{eq:plus_minus_basis_transform}. If we assume that the splitting is fast enough (less than a time scale of $d/c$ with $c$ speed of sound), patches at different locations are not correlated. Then, the variance of relative density is given by 
\begin{equation}
    \braket{\delta \hat{n}^-(z)\delta\hat{n}^-(z^\prime)} = \frac{2\braket{\Delta N^2}}{d^2}\delta_{zz^\prime} = \frac{N}{2}\frac{\delta_{zz^\prime}}{d}.
\end{equation}
Let $n_{\rm 1D}$ be the mean density after the splitting at each individual gas. The total number of atoms for a patch of size $d$ counting both wells is given by $N = 2n_{\rm 1D}d$. Hence, we get
\begin{equation}
    \braket{\delta\hat{n}^-(z)\delta\hat{n}^-(z^\prime)} = n_{\rm 1D}\frac{\delta_{zz^\prime}}{d} \approx n_{\rm 1D}\delta(z-z^\prime).
    \label{eq:real_space_corr}
\end{equation}
where the last approximation is valid for $d\rightarrow 0$ and $\delta(z-z^\prime)$ is the Dirac delta function. Note that our result differs by a factor of $1/2$ from Refs. \cite{gring2012relaxation, geiger2014local} due to $\sqrt{2}$ factor difference in our fields definition [see Eq. \eqref{eq:plus_minus_basis_transform}].

In momentum space, the real-space correlation \eqref{eq:real_space_corr} is equivalent to
\begin{equation}
\braket{\delta\hat{n}_{k}^-\delta\hat{n}_{q}^-} = n_{\rm 1D}\delta_{kq}, 
\end{equation}
which is the density-density part of Eq. \eqref{eq:init_corr} for unit squeezing $r = 1$. We assume that after coherent splitting, the relative sector is occupied by quantum fluctuation saturating the Heisenberg uncertainty principle
\begin{equation}
    \braket{\hat{\phi}_{k}^{-}\hat{\phi}_k^-}\braket{\delta\hat{n}_{k}^-\delta\hat{n}_k^-} \geq \frac{1}{4}.
\end{equation}
The real-space relative phase fluctuation is therefore estimated as
\begin{equation}
    \langle\hat{\phi}_-(z)\hat{\phi}_-(z^\prime)\rangle = \frac{1}{4n_{\rm 1D}}\delta(z-z^\prime),
\end{equation}
which is the real-space representation of the phase-phase correlation in Eq. \eqref{eq:init_corr} with $r = 1$. Meanwhile, all of the initial thermal energy of the gas gets stored in the symmetric sector
\begin{equation}
    \braket{\delta\hat{n}_k^+\delta\hat{n}_q^+} = \frac{\varepsilon_k^+n_{\rm 1D}}{E_k}(1+2\eta_k^+)\delta_{kq} \qquad \braket{\hat{\phi}_k^+\hat{\phi}_k^+} = \frac{E_k}{4n_{\rm 1D}\varepsilon_k^+}(1+2\eta_k^+)\delta_{kq},
\end{equation}
with $\eta_k^+ = \braket{(\hat{b}_k^+)^\dagger\hat{b}_k^+} = [\exp(\beta\varepsilon_k^+)-1]^{-1}$ being the thermal mean occupation number and $\beta$ is the inverse temperature.

\section{Derivation of logarithmic negativity and mutual information for coherently split gases}\label{appdxD}
Immediately after the coherent splitting, we assume thermal occupation in the symmetric sector and quantum noise in the antisymmetric sector [Eq. \eqref{eq:init_corr}]. The covariance matrix elements are now expressed as
\begin{equation}
    \braket{\hat{\phi}_k^a\hat{\phi}_k^b} = \begin{cases}
        &\frac{1}{8n_{\rm 1D}r^2}\left[\frac{r^2\varepsilon_k^+}{E_k}\left(1+2\eta_k^+\right)+1\right] \quad \text{if}\; a = b\\
        & \frac{1}{8n_{\rm 1D}r^2}\left[\frac{r^2\varepsilon_k^+}{E_k}\left(1+2\eta_k^+\right)-1\right] \quad \text{if}\; a \neq b,
    \end{cases}
\end{equation}
\begin{equation}
\braket{\delta\hat{n}_k^a\delta\hat{n}_k^b} = \begin{cases}
        &\frac{n_{\rm 1D}r^2}{2}\left[\frac{E_k}{\varepsilon_k^+}\frac{1}{r^2}(1+2\eta_k^+)+1\right]\quad \text{if}\; a = b\\
        & \frac{n_{\rm 1D}r^2}{2}\left[\frac{E_k}{\varepsilon_k^+}\frac{1}{r^2}(1+2\eta_k^+)-1\right] \quad \text{if}\; a \neq b,
    \end{cases}
\end{equation}
and all other elements are zeros. Since the covariance matrix $\Gamma$ has the same structure as in Eq. \eqref{eq:thermal_cov_elems_1} and \eqref{eq:thermal_cov_elems_2}, the calculation done in \ref{appdxB} can be extended to the split case. 

The symplectic eigenvalues for the partially transposed covariance matrix are calculated to be
\begin{equation}
     \nu_k^+\Big|_{\Gamma^{\top_B}} =\frac{1}{2}\sqrt{\mathcal{C}_{k,r}^{-1}\left(1+2\eta_k^+\right)} \qquad  \nu_k^-\Big|_{\Gamma^{\top_B}} =  \frac{1}{2}\sqrt{\mathcal{C}_{k,r}\left(1+2\eta_k^+\right)}.
     \label{eq:partial_tranposed_symp}
\end{equation}
where $\mathcal{C}_{k,r} = r^2\varepsilon_k^+/E_k$ is the squeezed spectral factor as defined in Eq. \eqref{eq:squeezed_spectral}.

Meanwhile, the symplectic eigenvalues of $\Gamma^{aa}$ ($a\in \{A,B\}$) are 
\begin{equation}
    \lambda_k\Big|_{\Gamma^{aa}} = \frac{1}{4}\sqrt{\left(1+2\eta_k+\mathcal{C}_{k,r}\right)\left(1+2\eta_k+\mathcal{C}_{k,r}^{-1}\right)},
    \label{eq:reduced_cov_symp}
\end{equation}
and the symplectic eigenvalues of the joint covariance matrix $\Gamma$ are
\begin{equation}
    \sigma_k^+\Big|_{\Gamma} = \frac{1}{2}(1+2\eta_k^+) \qquad \qquad \sigma_k^-\Big|_{\Gamma} = \frac{1}{2}.
    \label{eq:cov_symp}
\end{equation}
From the symplectic eigenvalues in Eqs. \eqref{eq:partial_tranposed_symp}-\eqref{eq:cov_symp} one obtains the expression for logarithmic negativity and mutual information after coherent splitting [Eq. \eqref{eq:log_neg_split} and Eq. \eqref{eq:split_mutual_info}]. 

The formula for the threshold temperature [Eq. \eqref{eq:split_threshold_temp}] is obtained by finding the root to
\begin{equation}
    \left(\nu_k^+-\frac{1}{2}\right)\left(\nu_k^--\frac{1}{2}\right) = 0.
    \label{eq:root}
\end{equation}
Recall that entanglement occurs when the partially transposed covariance matrix violates the Heisenberg uncertainty principle, i.e., either $\nu_k^+<1/2$ or $\nu_k^-<1/2$. In solving for the threshold temperature, we want to find a solution where at least one of them is exactly $1/2$. 

After substituting Eq. \eqref{eq:partial_tranposed_symp} to Eq. \eqref{eq:root}, the problem is reduced to solving
\begin{equation}
  4x^2 = (x^2-1)(\mathcal{C}_{k,r}+\mathcal{C}_{k,r}^{-1}+2)
    \label{eq:solve_crit}
\end{equation}
with $x = \exp(\varepsilon_k^+/(k_BT^*))$. The threshold temperature is obtained by $T^* = (k_B/\varepsilon_k^+)\ln x$ and $x$ is the solution to Eq. \eqref{eq:solve_crit}.

\bibliographystyle{quantum}
\bibliography{references.bib}
\end{document}